\documentclass[11pt,a4paper]{article}
\usepackage{jheppub,bm,booktabs,multirow}

\usepackage{color}
\usepackage{overpic}
\allowdisplaybreaks

\newcommand{\Stilde}[1]{\bm{\mathcal{ \widetilde{S}}}_{{#1}}} 
\newcommand{\Htilde}[1]{\bm{\mathcal{ \widetilde{H}}}_{{#1}}} 
 
\newcommand{\Ztilde}[1]{\bm{\mathcal{ \widetilde{Z}}}_{{#1}}} 
\usepackage[utf8x]{inputenc}

\makeatletter
\def\@fpheader{~}
\makeatother

\def\as{\alpha_s}
\def\e{\epsilon}
\def\nno{\nonumber}

\title{Factorization for the light-jet mass and hemisphere soft function}
\author[a]{Thomas Becher,}
\author[b]{Benjamin~D.~Pecjak,}
\author[a]{and Ding Yu Shao}
\affiliation[a]{Albert Einstein Center for Fundamental Physics, Institut f\"ur Theoretische Physik, Universit\"at Bern,
  Sidlerstrasse 5, CH-3012 Bern, Switzerland}
\affiliation[b]{Institute for Particle Physics Phenomenology, University of Durham, DH1 3LE Durham, United Kingdom}

\emailAdd{becher@itp.unibe.ch}
\emailAdd{ben.pecjak@durham.ac.uk}
\emailAdd{shao@itp.unibe.ch}

\date{\today}

\preprint{IPPP/16/84}

\abstract{Many collider observables suffer from non-global logarithms not captured by standard resummation techniques. Classic examples are the light-jet mass  event shape in the limit of small mass and the related hemisphere soft function. We derive factorization formulas for both of these and explicitly demonstrate that they capture all logarithms present at NNLO. These formulas achieve full scale separation and provide the basis for all-order resummations. A characteristic feature of non-global observables is that the soft radiation is driven by multi-Wilson-line operators, and the ones arising here map onto those relevant for the case of narrow-cone jet cross sections. Numerically, the contributions of non-global logarithms to resummed hemisphere-mass event shapes are sizeable.}

\begin{document}

\maketitle

\section{Non-global logarithms in hemisphere-mass observables}

Perturbative corrections to observables which involve a hierarchy of
scales are enhanced by logarithms of the scale ratios. Starting with
the pioneering work of Sudakov \cite{Sudakov:1954sw}, methods were
developed to resum such logarithmically enhanced corrections to all
orders. A crucial simplification is exponentiation, the statement that
the leading logarithms can be obtained from exponentiating the
leading-order correction to a process. Effective field theories provide a modern way to analyze
multi-scale problems. In these theories exponentiation is a consequence of the renormalization group
(RG). The logarithms are resummed by evolving Wilson
coefficients, which encode the physics associated with high scales,
down to lower scales and the leading-order solution of the RG
equation is an exponential.

Interestingly, this simple exponentiation property does not hold for
all observables. For example, if one considers interjet energy flow,
one finds that the relevant wide-angle soft radiation produces a very
intricate pattern of leading logarithms
\cite{Dasgupta:2002bw}. Instead of a simple linear evolution equation,
one needs to solve a complicated non-linear integral equation
to obtain the leading logarithms, the Banfi-Marchesini-Smye (BMS)
equation \cite{Banfi:2002hw}. Interjet energy flow is an example of a
non-global observable. Such observables are insensitive to radiation
in certain regions of phase space (the inside of the jets, for the case of the interjet energy flow) and the same complicated pattern of
``non-global'' logarithms  is present in all of them. Perhaps the
simplest quantity which suffers from such logarithms is the hemisphere
soft function, which is obtained by considering the radiation from two
Wilson lines in opposite directions. Allowing for large energy in one
hemisphere, but only a small amount in the other leads to non-global
logarithms. This soft function is also relevant in the context of the
light-jet mass event shape in $e^+e^-$ collisions, in which the complicated
pattern of logarithms was originally discovered
\cite{Dasgupta:2001sh}.

The BMS equation makes crucial use of the simple form of strongly
ordered gluon-emission amplitudes. Beyond leading logarithmic accuracy these
simplifications do not apply and it was therefore not clear how to
generalize the BMS equation to higher accuracy. In the past few years,
the problem of non-global logarithms has received renewed interest, in
particular in the context of Soft-Collinear Effective Theory (SCET)
\cite{Bauer:2000yr,Bauer:2001yt,Beneke:2002ph} (see
\cite{Becher:2014oda} for a review). Several papers have computed
hemisphere soft functions up to next-to-next-to-leading order (NNLO)  to obtain full
results for their non-global structure at this order
\cite{Kelley:2011ng,Hornig:2011iu,Kelley:2011aa,vonManteuffel:2013vja}. Furthermore,
by perturbatively expanding the BMS equation, the analytic form of the
leading-logarithmic terms up to five-loop order was extracted
\cite{Schwartz:2014wha,Khelifa-Kerfa:2015mma}. Using an efficient new
method to perform the angular integrations \cite{Caron-Huot:2016tzz},
this result has now been extended to 12\,(!) loops
\cite{CaronHuotUnpub}. In addition to these fixed-order
considerations, a method to approximately resum the non-global
logarithms was proposed \cite{Larkoski:2015zka,Neill:2015nya}. At
leading-logarithmic accuracy it reduces to an iterative solution of
the BMS equation \cite{Larkoski:2016zzc}.

In the recent papers \cite{Becher:2016mmh,Becher:2015hka}, two of us
have analyzed cone-jet cross sections and have derived factorization
theorems for the case where the outside energy is small. The
characteristic feature of these theorems is the presence of
multi-Wilson-line operators which describe the soft emissions from
energetic partons inside jets. In our effective-field-theory
framework, the non-global logarithms are obtained from an RG-evolution
equation which generalizes the BMS equation to arbitrary logarithmic
accuracy. The complicated structure arises because operators with an
arbitrary number of soft Wilson lines are present in the factorization
theorem. To obtain the large logarithms, one needs to exponentiate
an infinite-dimensional anomalous-dimension matrix, which, at
leading-logarithmic accuracy and large $N_c$, is equivalent to solving the BMS
equation. The exponentiation property mentioned earlier is thus present also for non-global logarithms, but takes a very complicated form. Our framework is closely related to the one proposed in
\cite{Caron-Huot:2015bja} and involves the same anomalous dimension, which was computed to two-loop order in that reference and has recently even been derived at three-loop accuracy in the planar
limit in $\mathcal{N}=4$ super Yang-Mills theory \cite{Caron-Huot:2016tzz}.

To make contact with the previous literature which has focused mostly
on the hemisphere soft function, it is important to analyze this
quantity using our framework. We do this in the present paper and at
the same time also derive a factorization theorem for the light-jet mass
event shape. To define this $e^+e^-$ event shape, one first introduces
the thrust axis $\vec{n}$ as the direction of maximum momentum
flow. More precisely, the unit vector $\vec{n}$ is chosen to maximize
the quantity $\sum_i |\vec{n} \cdot \vec{p}_i|$, where the sum runs
over all particles in the final state. The event shape thrust is
defined as this sum normalized to $Q$, where $Q$ is the center-of-mass
energy of the collision. The thrust axis splits each event into two
hemispheres, which can arbitrarily be labelled as ``left" and ``right",
and one can define additional event shapes by considering the
invariant masses $M_{L}$ and $M_{R}$ of the particles in the
hemispheres. Two commonly used event shapes are
\begin{align}
\text{heavy-jet mass:}\;\; \rho_h &= \frac{1}{Q^2} {\rm max}(M_L^2,M_R^2) \, ,\\
\text{light-jet mass:}\;\; \rho_\ell &=\frac{1}{Q^2}  {\rm min}(M_L^2,M_R^2)  \,.
\end{align}
In the limit where the jet masses become small, perturbative
corrections to these observables are logarithmically enhanced. For the
heavy-jet mass these logarithms have been resummed up to
next-to-next-to-next-to-leading logarithmic (N$^3$LL) accuracy
\cite{Chien:2010kc}, while only NLL predictions are available
for the light-jet mass $\rho_\ell$ \cite{Burby:2001uz,Dasgupta:2001sh}. The reason for the poor accuracy 
for $\rho_\ell$ was that it was not known how this non-global
observable factorizes in the limit of small $\rho_\ell$, while the
factorization is well known for the heavy-jet mass.

Due to left-right symmetry, the three possible scale hierarchies for
the hemisphere masses are a.) $M_L \sim M_R \ll Q$ , b.) $M_L \ll M_R
\ll Q$ and c.) $M_L \ll M_R \sim Q$. The relevant factorization
theorem for case a.) has the form \cite{Fleming:2007qr} 
\begin{equation}\label{eq:factMLMR}
\frac{d \sigma}{d M_L^2 d M_R^2}  = \sigma_0 H(Q^2) \int_0^\infty d\omega_L \int_0^\infty d\omega_R  \, J_q(M_L^2 - Q\,\omega_L)\, J_q(M_R^2 - Q\,\omega_R) \, S(\omega_L, \omega_R) \, ,
\end{equation}
where $\sigma_0$ is the Born level cross section. The hard function $H$
collects the virtual corrections to $\gamma^* \to q\bar{q}$ which are
known to three loops \cite{Becher:2006mr,Becher:2007ty}. The jet
function $J_q$ is the usual inclusive jet function in SCET, which is
known to two loops \cite{Becher:2006qw,Becher:2010pd}. The hemisphere
soft function $S(\omega_L, \omega_R)$ is a matrix element of Wilson
lines along the two jet directions and is also known at NNLO
\cite{Kelley:2011ng,Monni:2011gb,Hornig:2011iu}. This function
measures the contribution of the soft radiation to the hemisphere mass
in each hemisphere. Since the relevant anomalous dimensions are known
for all ingredients in \eqref{eq:factMLMR}, one can solve their RG
evolution equations to obtain N$^3$LL resummation for hierarchy a.)
which is the one relevant for the heavy-jet mass $\rho_h$.

However, the above theorem does not achieve resummation for case b.)
since for $\omega_L \ll \omega_R$ the soft function $S(\omega_L,
\omega_R) $ itself contains large logarithms of $\kappa =
\omega_L/\omega_R$, which are examples of non-global logarithms. To be
able to resum also these logarithms one must factorize the physics at
the two different soft scales $\omega_L$ and $\omega_R$. In the
context of the function $S(\omega_L, \omega_R) $, we will refer to
$\omega_R$ as the hard scale and $\omega_L$ the soft one. One of the main results
 of the present paper is that the hemisphere soft
function factorizes in the limit $\kappa\to 0$ as
\begin{align}
\label{sigbarefinal}
S(\omega_L, \omega_R)  &=  \sum_{m=0}^\infty \big\langle \bm{\mathcal{H}}^S_m(\{\underline{n}\},\omega_R) \otimes \bm{\mathcal{S}}_{m+1}(\{n,\underline{n}\},\omega_L) \big\rangle \,.
\end{align}
The hard functions $\bm{\mathcal{H}}^S_m$ are the squared amplitudes
for $m$-parton emissions from the two Wilson lines in the hemisphere soft function into the right hemisphere, integrated over
their energies but at fixed directions $\{\underline{n}\}=\{n_1,
\dots, n_m\}$, where the $n_i$'s are light-like vectors. The soft
functions $\bm{\mathcal{S}}_{m+1}$ consist of $m+2$ Wilson lines along
the directions $\{\underline{n}\}$ of the $m$ hard partons and the two
jets along $n^\mu = (1, \vec{n})$ and $\bar{n}^\mu = (1,
-\vec{n})$. Both of these are matrices in color space \cite{Catani:1996jh,Catani:1996vz}, and  
$\langle\dots \rangle$ indicates a sum over color indices.
The symbol $\otimes$ indicates that one has to integrate
over the $m$ directions of the emissions into the right
hemisphere. The form of the factorization theorem \eqref{sigbarefinal}
is basically the same as the one for wide-angle cone-jet cross
sections derived in \cite{Becher:2016mmh}. To see the connection, one
should view the right hemisphere as the inside of a jet which contains
hard particles with momenta $p^\mu \sim \omega_R$ and the left
hemisphere as the outside region where a veto on radiation is imposed
which constrains the momenta to $p^\mu \sim \omega_L$.

\begin{figure}[t!]
\begin{center}
\begin{tabular}{ccc}
\includegraphics[height=0.24\textwidth]{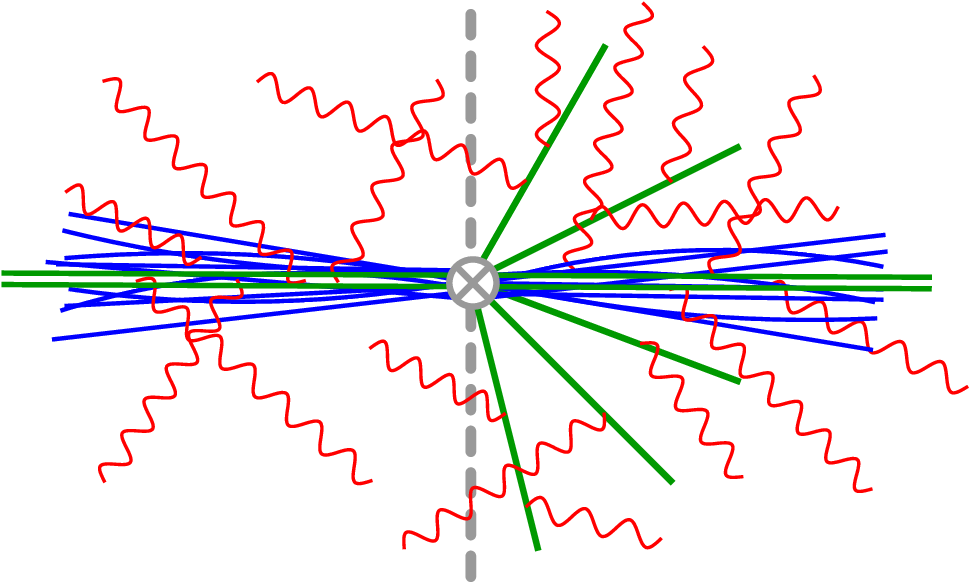} &\hspace{1cm} & 
\includegraphics[height=0.24\textwidth]{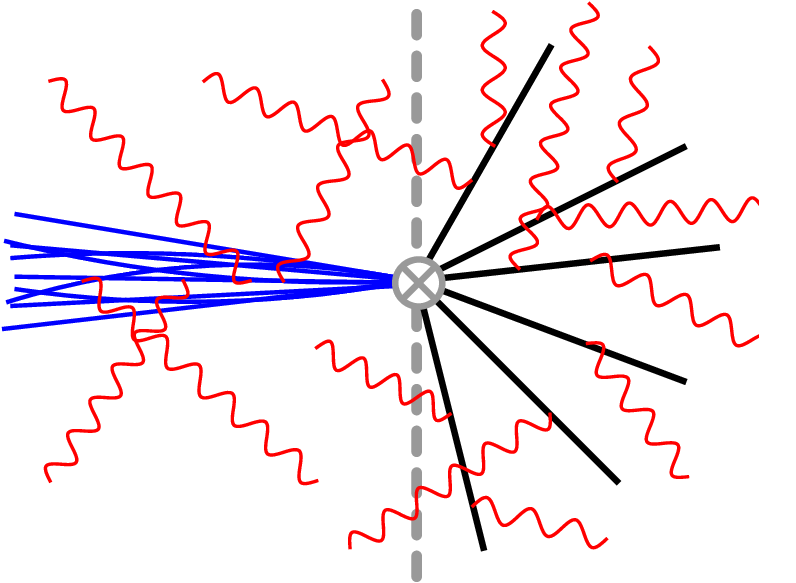} 
\end{tabular}
\end{center}
\vspace{-0.3cm}
\caption{Pictorial representation of the factorization theorems for the differential cross sections with respect to the hemisphere jet masses in the limit $M_L\ll M_R\ll Q$
 (left), and to the left-jet mass when $M_L\ll M_R \sim Q$ (right). Blue lines correspond to collinear partons inside the jet functions, the red lines represent soft emissions. The green lines in the left picture correspond to the hard part of the hemisphere soft function, while the black lines in the right picture correspond to hard emission into the right hemisphere.  \label{fig:fact}}
\end{figure}

Before analyzing the factorization formula \eqref{sigbarefinal} in
more detail and providing operator definitions for its ingredients, we
now turn to the light-jet mass $\rho_\ell$. Due to left-right symmetry
and its definition, $\rho_\ell$ is directly related to the left-jet
mass $\rho_L = M_L^2/Q^2$ according to
\begin{equation}\label{eq:lightjet}
\frac{d\sigma}{d\rho_\ell} =\left. 2 \frac{d\sigma}{d\rho_L} - \frac{d\sigma}{d\rho_h} \right|_{\rho_L=\rho_h=\rho_\ell}\,.
\end{equation}
Instead of the light-jet mass one can therefore equally well analyze
the factorization for $\rho_L$. If one only measures the left-jet
mass, the mass of the right jet will typically be large, so that scale
hierarchy c.) applies. We find that the cross section for the left-jet
mass factorizes as 
\begin{equation}\label{eq:factlightmass}
\frac{d \sigma}{d M_L^2}  = \sum_{i=q,\bar{q},g} \int_0^\infty d\omega_L\,  J_i(M_L^2 - Q\,\omega_L) \, \sum_{m=1}^\infty \big\langle \bm{\mathcal{H}}^i_m(\{\underline{n}\}, Q) \otimes \bm{\mathcal{S}}_m(\{\underline{n}\},\omega_L) \big\rangle \,.
\end{equation}
Since the unobserved radiation in the right hemisphere is typically hard, such that $p^\mu \sim Q$, we no longer encounter a jet function for this hemisphere, in contrast to the previous case \eqref{eq:factMLMR}. The hard functions also differ from the function $\bm{\mathcal{H}}^S_m$ encountered for the hemisphere soft functions. Rather than Wilson-line matrix elements as in \eqref{sigbarefinal}, the functions $\bm{\mathcal{H}}^i_m$ in this case are given by squared QCD amplitudes with a single parton of flavor $i$ in the left hemisphere propagating along the $\bar{n}$-direction and $m$ partons in the right hemisphere. The subsequent branchings of the hard parton on the left are described by the jet functions $J_i$. A graphical representation of the factorization theorems is shown in Figure \ref{fig:fact}.

Our paper is organized as follows. In the next section, we will flesh out the factorization formulas for the hemisphere soft function and for the light-jet mass event shape and discuss their derivation, which can be obtained following similar steps as in \cite{Becher:2016mmh}. The soft functions in these theorems can be related to the coft functions computed in that reference so that the only new ingredients to our factorization formulas are the hard functions. After computing these in Section \ref{sec:hemishpere} up to $\mathcal{O}(\alpha_s^2)$, we verify that we reproduce the known NNLO result for the hemisphere soft function in the limit $\omega_L \to 0$. Next, we analyze the light-jet mass distribution in Section \ref{sec:lightjet} and compare to the numerical fixed-order result for this quantity. In Section \ref{sec:NLL} we use the known result for the leading non-global logarithms in the 
hemisphere soft function to obtain numerical results for the light-jet mass at NLL accuracy. In Section \ref{sec:conclusions} we discuss the necessary steps to perform higher-order resummation for this event shape and conclude.

\section{Factorization\label{sec:fact}}

The derivation of the factorization formula follows the same steps in both cases and is similar to the one relevant for wide-angle cone-jet cross sections presented in \cite{Becher:2016mmh}. We will first sketch the derivations of the theorems and specify the ingredients. We then relate the soft functions to the ones which arise in the case of the narrow-cone jet cross sections. Due to this relation, we can use the results \cite{Becher:2016mmh} for these and only the hard functions need to be computed.

\subsection{Hemisphere soft function}

The hemisphere soft function describes radiation originating from a quark and an anti-quark along the directions $n$ and $\bar{n}$ of the two jets. Their soft radiation is described by Wilson lines. The one generated by the outgoing quark along the $n$ direction is
\begin{align}
\label{eq:Sn}
S(n)=\, \bm{P} \exp\left( ig_s \int_0^\infty \!ds \,n\cdot A^a(s n) t^a \right)  ,
\end{align}
and the soft function is defined as
\begin{align}
\label{eq:SoftFunction}
S(\omega_L, \omega_R) = \frac{1}{N_c} \sum_{X} {\rm Tr} \langle 0 | S(\bar{n}) S^\dag( n) | X \rangle \langle X | S(n) S^\dag(\bar{n}) | 0 \rangle \delta(\omega_R - n\cdot P_R) \,  \delta(\omega_L - \bar{n}\cdot P_L) \,,
\end{align}
where the trace is over color indices. We call the hemisphere which
contains the thrust vector the right hemisphere. The right-moving
particles therefore have $\bar{n}\cdot p > n\cdot p$ and $P_{R(L)}$ is
the total momentum in the right (left) hemisphere. Usually, the
function $S(\omega_L, \omega_R)$ is defined in terms of the soft gluon
field in SCET. However, the soft SCET Lagrangian is equivalent to the
full QCD one so for our discussion we will consider
\eqref{eq:SoftFunction} as a matrix element in QCD. In the asymmetric
case $\omega_L \ll \omega _R$ the function $S(\omega_L, \omega_R)$
develops large, non-global logarithms (NGLs) in the ratio $\kappa
\equiv \omega_L/\omega_R \ll 1$. It is these logarithms which we seek
to resum using effective-field-theory methods.

Before constructing the appropriate effective theory, it is useful to
study the structure of NGLs in the matrix element
\eqref{eq:SoftFunction} perturbatively.  Clearly, one method is to
calculate the hemisphere soft function at a given order in
perturbation theory, and then take the limit $\kappa \to 0$ in the
final result.  This was the approach taken in the
NNLO calculations of
\cite{Hornig:2011iu, Kelley:2011ng}, and the obvious benefit of such a
computation is that it provides the hemisphere soft function for any
value of $\kappa$.  On the other hand, if one is interested only in
NGLs appearing in the limit $\kappa \to 0$, it is much simpler to
obtain results by expanding the phase-space integrals appearing in the
hemisphere soft function using the method of regions
\cite{Beneke:1997zp}.  Indeed, in a first step we have used this
method to reproduce the NNLO fixed-order calculations in the
non-global limit.  The factorization results discussed below can be
viewed as a translation of this diagrammatic approach into the
language of effective field theory.

We find that two momentum regions are needed for the leading-power
diagrammatic expansion in the limit $\kappa \to 0$. Defining the
light-cone components of an arbitrary vector $p$ as $(n\cdot p,
\bar{n}\cdot p ,p_\perp)$, these regions are specified by the scalings
\begin{equation}
\label{eq:MomScalings}
\begin{aligned}
  \mbox{hard:}\quad  p_{h} &\sim \omega_R \,(1,1, 1)  \,,  \\
  \mbox{soft:}\quad     p_{ s} &\sim  \omega_R  \,(\kappa, \kappa, \kappa) \,.
\end{aligned}
\end{equation}
The homogeneous scaling of the momentum components arises because the
soft and hard radiation covers a wide angular range so that no
specific direction is singled out. The expansion of individual
diagrams also receives contributions from a left-collinear mode
scaling as $\omega_R (1,\kappa,\sqrt{\kappa})$.  However, in the sum
of all diagrams these collinear contributions vanish, and in Appendix
\ref{sec:Jetfun} we present an all-orders proof of this result, based
on the invariance of Wilson lines under rescalings of the reference
vector.

A non-trivial interplay between contributions of the two regions is
responsible for the structure of NGLs in the hemisphere soft
function. By NGLs, we mean contributions which cannot be written as a
naive product of two component functions depending on $\omega_L$ and
$\omega_R$ only.\footnote{The exact definition of NGLs is ambiguous;
  we consider several possibilities below in the discussion following
  (\ref{eq:Sng}).}  An NLO analysis does not reveal the presence of
NGLs, since the NLO result is the sum of the identical contributions
of a single hard emission into the right hemisphere and a single soft
emission into the left hemisphere, which can always be written as the
product of identical one-scale functions for the hard and soft
regions.  At NNLO, on the other hand, it is possible for a virtual gluon to split into two particles flying into different hemispheres, and it is obvious
that a simple product structure is insufficient to describe these
contributions since they have a different
color structure.  Two types of opposite-hemisphere configurations are relevant. 
The first  involves a soft gluon in the left hemisphere and a
hard gluon in the right hemisphere and gives rise to double and
single NGLs. The second involves one soft gluon in each hemisphere.
Such a configuration is not possible for hard radiation, because a hard emission into the left hemisphere
would violate the scaling $\omega_L\ll \omega_R$.  This asymmetry
between double-hard and double-soft contributions generates the
remaining single NGLs needed to reproduce the known NNLO result in the
$\kappa \to 0$ limit.

The effective field theory appropriate for describing the situation
above has recently been developed in \cite{Becher:2015hka,
  Becher:2016mmh}. The basic observation of these papers was that each
of the hard partons generates a soft Wilson line along its direction,
so even though hard and soft contributions factorize in (\ref{sigbarefinal}), new hard and soft functions
  appear at each order in perturbation theory.  To obtain the
operators in the low-energy effective theory, one therefore first
considers a kinematic configuration with $m$ hard partons along fixed
directions and then introduces a soft Wilson line for each of
them. The amplitudes for the emissions of $m$ hard partons with
momenta $\{\underline{p}\}=\left\{p_1,\cdots, p_m \right\}$ from the
two Wilson lines in \eqref{eq:SoftFunction} take the form
\begin{equation}\label{hardAmpS}
| \mathcal{M}^S_m(\{\underline{p} \}) \rangle = \langle \{\underline{p}\} | S(n) S^\dag(\bar{n}) | 0 \rangle\,.
\end{equation}
Note that on the left-hand side of the above equation we use the
color-space formalism of \cite{Catani:1996jh,Catani:1996vz} in which
the amplitude $| \mathcal{M}^S_m(\{\underline{p})\} \rangle$ is a
vector in the color space of the $m$ partons.  However, on the
right-hand side the color indices of the $m$ partons are suppressed
and the bra-ket notation denotes states in the Hilbert space.  The
superscript $S$ indicates that the amplitude $\mathcal{M}^S_m$ is obtained from the
Wilson line matrix element.

A general soft Wilson line along the light-like direction $n_i \propto p_i$ is defined in analogy with (\ref{eq:Sn}) as
\begin{equation}\label{eq:Si}
\bm{S}_i(n_i) = {\bf P} \exp\left( ig_s \int_0^{\infty}\!ds\,
   n_i\cdot A_s^a( s n_i)\,\bm{T}_i^a \right) ,
\end{equation}
where the color matrices for the representation of the underlying
particle $i$ are denoted by $\bm{T}_i^a$. On the amplitude level, the
soft radiation from the two original Wilson lines and the additional
hard partons is obtained from the Wilson-line operator
\begin{equation}\label{eq:softfact}
\bm{S}_a(\bar{n})\,\bm{S}_b(n)\,\bm{S}_1(n_1) \dots {\bm S}_m(n_m) | \mathcal{M}^S_m(\{\underline{p} \} \rangle \,,
\end{equation}
where $\bm{S}_a(\bar n)$ and $\bm{S}_b( n)$ are the anti-quark and quark Wilson lines present in the original definition \eqref{eq:SoftFunction}. A derivation of the formula \eqref{eq:softfact} from SCET was given in \cite{Becher:2016mmh}. 

To obtain the factorized result for the cross section we need to
square the factorized amplitude \eqref{eq:softfact}, integrate over
the energies and directions of the hard partons, and add up the
contributions from different multiplicities of hard partons. Doing so,
we obtain the factorization formula \eqref{sigbarefinal} for the
hemisphere soft function in the limit $\kappa \to 0$. The definitions
of the hard functions in this formula read
\begin{align}
\bm{\mathcal{H}}^S_{m}(\{\underline{n}\},\omega_R) 
  =  \prod_{i=1}^m &\int \! \frac{dE_i \,E_i^{d-3} }{(2\pi)^{d-2}} \,  | \mathcal{M}^S_m(\{\underline{p}\}) \rangle \langle \mathcal{M}^S_m(\{\underline{p}\}) |  \,
   \delta(\omega_R - n \cdot P_R)\, {\Theta }_{R}\!\left(\left\{\underline{p}\right\}\right) ,
\end{align}
where $d$ is the number of spacetime dimensions. The theta
function ${\Theta }_{R}$ ensures that all hard partons are inside the
right hemisphere so that $P_R$ is simply the total hard momentum.  Note that the directions of the hard partons are fixed. The integral over the directions is performed after multiplication with the soft function, which for $m$ additional hard partons is obtained from
squaring the Wilson-line operator matrix elements
\begin{align}\label{eq:SmDef}
   \bm{\mathcal{S}}_{m+1}(\{n, \underline{n}\},\omega_L) = \int\limits_{X_s}\hspace{-0.58cm} \sum\, &
 \langle 0 | \,\bm{S}^\dag_a(\bar{n})\,\bm{S}^\dag_b( n)\,\bm{S}^\dag_1(n_1) \dots {\bm S}^\dag_m(n_m)\,  | X_s \rangle \nonumber \\
  & \times \langle X_s| \,\bm{S}_a(\bar n)\,\bm{S}_b( n)\,\bm{S}_1(n_1) \dots {\bm S}_m(n_m)\, |0\rangle \, \delta(\omega_L - \bar{n}\cdot P_L)\,.    
\end{align}
Note that the soft partons can be in either hemisphere. The ones in
the left hemisphere contribute to $\omega_L$, but the ones in the
right hemisphere are not constrained because their contribution to
$\omega_R$ is negligible compared to the hard partons. The strict
expansion of the phase-space measure is crucial to achieve the desired
factorization of scales and to avoid double counting of the
contributions from different momentum regions.

\subsection{Left-jet mass}

The factorization for the left-jet mass distribution is rather similar to that for the hemisphere soft function, but the expansion parameter is $\lambda = \omega_L/Q$ and the relevant momentum scalings are
\begin{equation}
\label{eq:MomScalingsLeft}
\begin{aligned}
  \mbox{hard:}\quad  p_{h} &\sim Q \,(1,1, 1)  \,,  \\
  \mbox{soft:}\quad     p_{ s} &\sim  Q  \,(\lambda, \lambda, \lambda) \,,\\
   \mbox{collinear:}\quad  p_{c} &\sim  Q  \,(1, \lambda, \sqrt{\lambda}) \, .
\end{aligned}
\end{equation}

To derive the factorization theorem (\ref{eq:factlightmass}) and obtain the hard functions
$\bm{\mathcal{H}}^i_m(\{\underline{n}\}, Q)$, one can first match onto
a version of SCET with a collinear field along the $\bar{n}$-direction as
well as $m$ additional collinear fields along directions in the right
hemisphere. Then one performs the usual decoupling transformation on
the collinear fields \cite{Bauer:2001yt}, which gives rise to the relevant
soft multi-Wilson-line operator. Finally one takes the matrix element
where there is a single hard parton along each of the $m$ directions
in the right hemisphere, and a jet of partons along the
$\bar{n}$-direction on the left. This yields the hard functions
$\bm{\mathcal{H}}^i_m(\{\underline{n}\}, Q)$ together with the jet
function $J_i$. We refrain from going over this derivation in more
detail since it involves, up to obvious modifications, exactly the
same steps as the ones detailed for the wide-angle jet cross section
in \cite{Becher:2016mmh}.

The explicit definition of the hard functions for the the decay of
  a virtual photon into a final state with $m$ particles in the right
  hemisphere is 
\begin{align}
\label{eq:LightJetHard}
\bm{\mathcal{H}}^i_{m}(\{\underline{n}\},\omega_R) 
  = \frac{1}{2Q} \prod_{j=1}^m &\int \! \frac{dE_j \,E_j^{d-3} }{(2\pi)^{d-2}} \,  | \mathcal{M}^{i}_{m+1}(\{p_0,\underline{p}\}) \rangle \langle \mathcal{M}^{i}_{m+1}(\{p_0,\underline{p}\}) |  
  \nonumber \\
&\times    \delta(\omega_R - n \cdot P_R)\, {\Theta }_{R}\!\left(\left\{\underline{p}\right\}\right)
   (2\pi)^d \,\delta(Q-E_{\rm tot}) \,\delta^{(d-1)}(\vec{p}_{\rm tot}) \,,
\end{align}
where $p_0^\mu= Q\, \bar{n}^\mu/2$ is the momentum of the single
hard parton of flavor $i\in\{q,\bar{q},g\}$ in the left hemisphere, and the amplitudes $|
\mathcal{M}^i_{m+1}(\{p_0,\underline{p}\}) \rangle$ are standard QCD
amplitudes for the decay of the virtual photon into $(m+1)$
partons. The associated soft function is
\begin{align}\label{eq:SmDefLight}
   \bm{\mathcal{S}}_m(\{ \underline{n}\},\omega_L) = \int\limits_{X_s}\hspace{-0.58cm} \sum\, &
 \langle 0 | \,\bm{S}^\dag_0(\bar{n})\,\bm{S}^\dag_1(n_1) \dots {\bm S}^\dag_m(n_m)\,  | X_s \rangle \nonumber \\
  & \times \langle X_s| \,\bm{S}_0(\bar n) \,\bm{S}_1(n_1) \dots {\bm S}_m(n_m)\, |0\rangle \, \delta(\omega_L - \bar{n}\cdot P_L)\,.    
\end{align}
This is exactly the same matrix element as \eqref{eq:SmDef} up to the fact that only the direction of the first Wilson line is fixed, as opposed to the case of the hemisphere soft function, where the first two, along the $\bar{n}$ and $n$ directions, are kept fixed. We can thus get the one in \eqref{eq:SmDef} by taking the result for \eqref{eq:SmDefLight} and setting the 
reference vector of the second Wilson line to the $n$ direction. 

Furthermore, almost the same matrix element as \eqref{eq:SmDefLight}
has arisen in the context of narrow-cone jet cross sections. In that case, 
the Wilson line structure is associated with coft emissions
which are simultaneously collinear and soft. Rather than a hemisphere
constraint, the coft functions involve a constraint on out-of-jet
radiation of the form $Q\beta > \bar{n}\cdot p_{{\rm\,out}}$ and a
particle is outside the right jet if $n\cdot p >
\delta^2\,\bar{n}\cdot p$. If we set $\delta=1$ and replace $Q\beta
\to \omega_L$, the coft functions are mapped onto the left hemisphere
(up to the fact that we impose the constraint as a $\delta$-function
instead of an upper limit). Since Wilson lines are invariant under a
rescaling of the reference vector, the transformation maps the coft
Wilson line matrix elements directly onto the soft functions
\eqref{eq:SmDefLight} and we can use the results of
\cite{Becher:2015hka,Becher:2016mmh}.

\section{Hemisphere soft function at NNLO\label{sec:hemishpere}}

In this section we demonstrate how our factorization formula can be
used to reproduce the results for the hemisphere soft function at NNLO
in perturbation theory in the asymmetric limit $\omega_L \ll
\omega_R$.  In the following, it will be convenient to work in Laplace
space, where the convolutions in the factorization formulas
\eqref{eq:factMLMR} and \eqref{eq:factlightmass} turn into products.
We define the renormalized, Laplace-transformed soft function as
\begin{align}
\label{eq:LaplaceS}
\tilde{s}(\tau_L,\tau_R,\mu)= 
\int_0^{\infty}  d\omega_L \int_0^{\infty} d\omega_R\, e^{-\omega_L/(\tau_L e^{\gamma_E})}
e^{-\omega_R/(\tau_R e^{\gamma_E})}
S(\omega_L, \omega_R, \mu) \,.
\end{align}
Whereas the soft function is a distribution in the arguments $\omega_{L,R}$, the Laplace-transformed 
soft function is a regular function of its arguments. 
The renormalized soft function in Laplace space is obtained from the bare one 
through multiplication by a UV renormalization factor.  We write the relation between the bare 
and renormalized functions as 
\begin{align}\label{eq:softren}
\tilde{s}(\tau_L,\tau_R,\mu)=   
\tilde{Z}_S(\tau_L, \tau_R ,\epsilon, \mu) \tilde{s}(\tau_L,\tau_R,\epsilon) \,.
\end{align}  
The notation, used throughout the paper, is such that bare and renormalized 
functions are distinguished through their last argument,  which is $\mu$ for renormalized
functions and $\epsilon$ for bare ones, where the dimensional regulator is $\epsilon=(4-d)/2$.  
On the other hand, in generic expressions such as (\ref{sigbarefinal}), we drop the dependence on $\mu$ or $\epsilon$ to 
indicate that the equations can refer equally well to bare or renormalized quantities.
The form and explicit results for the renormalization factor
$\tilde{Z}_S$ are well known -- we collect some of the expressions we need in the analysis below in 
Appendix~\ref{sec:Barefun}.  

We now show how to reproduce the NNLO results of \cite{Hornig:2011iu, Kelley:2011ng} for 
the hemisphere soft function  using the factorization formalism from the previous section. 
We first define the Laplace-transformed component functions as 
\begin{align}
\label{eq:LaplaceHm}
\bm{\mathcal{\widetilde{H}}}^S_m(\{\underline{n}\}, \tau_R) & =\int_0^\infty d\omega_R \, e^{-\omega_R/(\tau_R e^{\gamma_E})}\bm{\mathcal{H}}^S_m(\{\underline{n}\},\omega_R) \, 
\end{align}
and
\begin{align}
\label{eq:LaplaceSm}
\bm{\mathcal{\widetilde{S}}}_m(\{\underline{n}\}, \tau_L) & =\int_0^\infty d\omega_L\, e^{-\omega_L/(\tau_L e^{\gamma_E})}\bm{\mathcal{S}}_m(\{\underline{n}\},\omega_L) \, .
\end{align}
The functions with different numbers of hard partons mix under renormalization. Following \cite{Becher:2016mmh},  we define the renormalized hard functions according to 
\begin{align}
\label{eq:HmZfac}
 \bm{\mathcal{ \widetilde{H}}}^S_m(\{\underline{n} \} ,\tau_R,\epsilon)
 = \sum_{l=0}^{m} \bm{\mathcal{ \widetilde{H}}}^S_l(\{\underline{n} \} ,\tau_R,\mu) \,
  \bm{\mathcal{ \widetilde{Z}}}_{lm}(\{\underline{n} \} ,\tau_R, \e,\mu)  \,.
\end{align}
This equation states that lower-multiplicity hard functions absorb
some of the divergences of the higher-point functions. This is
familiar from fixed-order computations, where virtual corrections to
lower-point amplitudes need to be combined with real-emission
contributions.

Combined with the fact that the UV divergences for the hemisphere soft
function are removed by the renormalization factor $\tilde{Z}_S$, the renormalized soft functions can be written
as
\begin{align}
\label{eq:SmZfac}
 \bm{\mathcal{ \widetilde{S}}}_{l+1}(\{\underline{n} \} ,\tau_L,\mu)=\sum_{m=l}^{\infty}
\left[\tilde{Z}_S(\tau_L,\tau_R,\epsilon,\mu) \,\bm{\mathcal{ \widetilde{Z}}}_{lm}(\{\underline{n} \} ,\tau_R, \e,\mu)\right]  \hat{\otimes} \,
 \bm{\mathcal{ \widetilde{S}}}_{m+1}(\{\underline{n} \} ,\tau_L,\epsilon) \,.
 \end{align}
The peculiar index structure arises because in the factorization theorem \eqref{sigbarefinal} for the hemisphere soft function, the hard function $\bm{\mathcal{ \widetilde{H}}}^S_{m}$ multiplies $\bm{\mathcal{ \widetilde{S}}}_{m+1}$. This relation has several non-trivial features. First of all, it implies that higher-multiplicity soft functions enter the renormalization of lower-multiplicity ones. The higher-$m$ functions depend on additional directions which need to be integrated over. This integral over unresolved directions is indicated by the symbol $\hat{\otimes}$. Both $\tilde{Z}_S$ and the $\bm{\mathcal{ \widetilde{Z}}}_{lm}$ depend on the hard scale $\tau_R$. It is a non-trivial cross check on our results that the renormalized soft function depends only on 
$\tau_L$, as it must.

The Laplace-transformed hemisphere soft function satisfies a factorization formula of
the same form as \eqref{sigbarefinal}.  In order to verify it to NNLO,
we first define expansion coefficients of the bare and renormalized functions as 
\begin{align}
\label{eq:expansion}
\tilde{s}(\tau_L,\tau_R,\e) & = 
\sum_{n=0}^\infty \left(\frac{\alpha_0}{4\pi}\right)^n \tilde{s}^{(n)}(\tau_L,\tau_R,\e) \, , \\
\tilde{s}(\tau_L,\tau_R,\mu) & = 
\sum_{n=0}^\infty \left(\frac{\alpha_s}{4\pi}\right)^n \tilde{s}^{(n)}(\tau_L,\tau_R,\mu) \, ,
\end{align} 
and similarly for the component functions $\bm{\mathcal{\widetilde{H}}}^S_m$ and $\bm{\mathcal{\widetilde{S}}}_m$.  Our definitions are such that
bare coupling constant in $d$-dimensions is written as
$\alpha_0 \tilde{\mu}^{2\epsilon}$, where $\tilde{\mu}^2=\mu^2 e^{\gamma_E}/(4\pi)$
is chosen to obtain results in the $\overline{\rm
MS}$ scheme. The renormalized coupling constant
$\alpha_s\equiv \alpha_s(\mu)$ is related to the dimensionless coupling
constant  $\alpha_0$ as $\alpha_s = Z^{-1}_\alpha \alpha_0$, where
\begin{align}
Z_\alpha =1- \frac{\alpha_s}{4\pi} \frac{\beta_0}{\epsilon} + \dots \, ; 
\qquad \beta_0 = \frac{11}{3}C_A - \frac{4}{3} T_F n_f \,.
\end{align}

Writing out the contributions to the factorization theorem \eqref{sigbarefinal} to first order, we obtain
\begin{align}
\label{eq:NLOfact}
\tilde{s}^{(1)}(\tau_L,\tau_R) =&\big\langle  \bm{\mathcal{\widetilde{H}}}^{S(0)}_0(\tau_R ) \,\bm{\mathcal{\widetilde{S}}}_1^{(1)}(\{n\},\tau_L) \big\rangle + \big\langle  \bm{\mathcal{\widetilde{H}}}^{S(1)}_0(\tau_R ) \, \bm{\mathcal{\widetilde{S}}}_1^{(0)}(\{n\},\tau_L) \big\rangle  \nonumber \\
& + \big\langle \bm{\mathcal{\widetilde{H}}}^{S(1)}_1(\{n_1\},\tau_R) \otimes \bm{\mathcal{\widetilde{S}}}_2^{(0)}(\{n,n_1\},\tau_L) \big\rangle \,,
\end{align}
where we have made explicit that the two terms on the first line have
no angular dependence, so that the convolution of functions reduces to
simple product.  Higher-multiplicity terms do not arise since the hard
functions are suppressed, $\bm{\mathcal{\widetilde{H}}}^S_{m}\sim
\alpha_s^m$. The formula simplifies further after noting that
perturbative corrections to the zero-emission hard function are
scaleless and vanish in dimensional regularization, so that
$\bm{\mathcal{\widetilde{H}}}^S_0(\tau_R,\epsilon )=\bm{1}$.
Furthermore the leading order soft functions
$\bm{\mathcal{\widetilde{S}}}_m^{(0)}=\bm{1}$ are trivial since the
Wilson lines reduce to unit matrices at leading order. Suppressing the
dependence on the arguments, the one-loop result reads
\begin{align}
\label{eq:NLOfactS}
\tilde{s}^{(1)}(\tau_L,\tau_R) =&\big\langle \bm{\mathcal{\widetilde{S}}}_1^{(1)} \big\rangle  + \big\langle \bm{\mathcal{\widetilde{H}}}^{S(1)}_1 \otimes \bm{1} \big\rangle \,.
\end{align}
Applying the same simplifications, the NNLO coefficient reads
\begin{align}
\label{eq:NNLOfact}
\tilde{s}^{(2)}(\tau_L,\tau_R)  = \langle \bm{\mathcal{ \widetilde{S}}}_1^{(2)} \rangle  
+  \langle \bm{\mathcal{\widetilde{H}}}_1^{S(1)}  \otimes   \bm{\mathcal{\widetilde{S}}}_2^{(1)}   \rangle
+ \langle \bm{\mathcal{ \widetilde{H}}}_1^{S(2)}  \otimes {\bm 1}   \rangle
+  \langle \bm{\mathcal{ \widetilde{H}}}_2^{S(2)}  \otimes {\bm 1}   \rangle \,.
\end{align}
In the following, we give explicit results for the ingredients in these two formulas. We can evaluate equations \eqref{eq:NLOfactS} and \eqref{eq:NNLOfact} using bare ingredients or renormalized ones. In the main text, we will work with renormalized quantities, but in Appendix~\ref{sec:Barefun} we repeat the computation using bare ones.

\subsection{Soft functions\label{sec:soft}}

As we stressed at the end of Section \ref{sec:fact}, the soft functions are trivially related to the coft functions $\bm{\mathcal{U}}_m$ relevant for narrow-jet cross sections defined in \cite{Becher:2015hka,Becher:2016mmh}. Indeed, after setting the cone-angle parameter $\delta=1$, the soft function for the left-jet mass \eqref{eq:SmDefLight} is identical to the coft function
\begin{equation}\label{eq:relCoft}
\bm{\mathcal{\widetilde{S}}}_m(\{\underline{n}\}, \tau_L) =\bm{\mathcal{\widetilde{U}}}_m(\{\underline{n}\},\tau_L) \,.
\end{equation}
As discussed after \eqref{eq:SmDefLight}, for the case of the
hemisphere soft function the first reference vector must be set equal
to $n^\mu$, see \eqref{eq:SmDef}, because the Wilson line along the
$n$-direction is present in the original hemisphere soft function
\eqref{eq:SoftFunction} and only the remaining $(m-1)$ Wilson lines
arise from hard partons. To be able to use our results in both cases,
we will give results for the left-jet mass case.

The one-loop soft function is a sum over dipoles
\begin{align}\label{eq:SmNLO}
   \bm{\mathcal{S}}_m(\{\underline{n}\},\omega_L,\e) =  \bm{1} 
   - g_s^2\,\tilde{\mu}^{2\e} \sum_{(ij)}\,\bm{T}_i\cdot\bm{T}_j 
  \int\! &\frac{d^{d-1}k}{(2\pi)^{d-1} 2E_k}\,\frac{n_i\cdot n_j}{n_i\cdot k\,n_j\cdot k}\, \nonumber\\
    & \theta(n\cdot k-\bar{n}\cdot k) \delta(\omega_L - \bar{n}\cdot k)
    + \dots \,,
\end{align}
where the summation of $(ij)$ goes over all unordered pairs, and we can restrict the soft emission to the left hemisphere because the contribution from the right hemisphere is a scaleless integral. 

It is useful to separate out the dipoles involving the left-Wilson line $\bm{S}_0(\bar{n})$ and write the one-loop coefficient of the function in Laplace space in the general form
\begin{align}\label{eq:SmNLOE}
   \bm{\mathcal{\widetilde{S}}}^{(1)}_m&(\{\underline{n}\},\tau_L) =  
  - \sum_{i}\,\bm{T}_0\cdot\bm{T}_i \, u(\hat{\theta}_i,\tau_L)  - \frac{1}{2} \sum_{[ij]}\,\bm{T}_i\cdot\bm{T}_j \, v(\hat{\theta}_i,\hat{\theta}_j, \phi_i - \phi_j,\tau_L) \, ,
  \end{align}
where the summation of $[ij]$ goes over all unordered pairs with $i,j\neq 0$. Here $\phi_i$ is the angle of the $n_i$ in the plane transverse to the thrust direction and
\begin{equation}\label{hatTheta}
\hat{\theta}_i = \sqrt{\frac{n\cdot n_i}{\bar{n}\cdot n_i}} = \tan\!\left(\frac{\theta_i}{2}\right) 
\end{equation}
parameterizes the angle with respect to the thrust axis. Since the
terms in the first sum depend only on a single reference vector $n_i$,
the coefficient $u(\hat{\theta}_i,\tau_L)$ is a function of the
corresponding angle. The result for the renormalized coefficient
functions can be obtained from the results for the coft function
$\bm{\mathcal{\widetilde{U}}}_2$ given in \cite{Becher:2016mmh}. We
find
\begin{align}
u(\hat{\theta}_1,\tau_L,\mu) & =-4 \ln^2 \Big(\frac{\tau_L}{\mu}\Big)-4 \ln  \Big( \frac{\tau_L}{\mu}\Big)\, \ln \left(1-\hat{\theta
   }_1^2\right) +  f_0\left(\hat{\theta }_1\right)-\frac{\pi ^2}{2}\,, \\
v(\hat{\theta}_1,\hat{\theta}_2, \Delta\phi ,\tau_L,\mu) & =2 g_0\!\left(\hat{\theta}_1,\hat{\theta}_2, \Delta\phi \right)+f_0\!\left(\hat{\theta}_1\right)-f_0\left(\hat{\theta}_2\right) \nonumber\\
& \hspace{3cm}+4  \ln\!\left(\frac{1+\hat{\theta}_1^2 \hat{\theta}_2^2-2 \hat{\theta}_1 \hat{\theta}_2 \cos\Delta\phi}{(1-\hat{\theta}_1^2) (1-\hat{\theta}_2^2)}\right) \ln \frac{\tau_L}{\mu}\,.
\end{align}
The function $u$ involves double logarithms due to a collinear singularity from the region where the emission is collinear to $\bar{n}$. The function $v$ on the other hand, describes an exchange between Wilson lines in the right hemisphere. Since the gluon is emitted to the left, this function does not suffer from a collinear singularity.  The auxiliary functions $f_0$ and $g_0$ were given in \cite{Becher:2016mmh} and read
\begin{align}\label{eq:f0g0}
   f_0(\hat\theta_1) &= - 2\ln^2(1-\hat\theta_1^2) - 2{\rm Li}_2(\hat\theta_1^2) \,, \\[2mm]
   g_0(\hat\theta_1,\hat\theta_2,\pi) 
   &= - \ln^2(1-\hat\theta_1^2) - 3\ln^2(1-\hat\theta_2^2)
    + 2 \left[ \ln(1-\hat\theta_1^2) + \ln(1-\hat\theta_2^2) \right] 
    \ln(1+\hat\theta_1\hat\theta_2) \nonumber\\
   &\quad - 2 \, {\rm Li}_2(\hat\theta_2^2) + 2 \, {\rm Li}_2(-\hat\theta_1\hat\theta_2) 
    - 2\,{\rm Li}_2\bigg( \! - \frac{\hat\theta_1^2+\hat\theta_1\hat\theta_2}{1-\hat\theta_1^2} \bigg) 
    - 2\,{\rm Li}_2\bigg( \! - \frac{\hat\theta_2^2+\hat\theta_1\hat\theta_2}{1-\hat\theta_2^2} \bigg) \, . \nonumber
\end{align}
For the function $\bm{\mathcal{\widetilde{S}}}_2$, it is sufficient to consider the case $\Delta \phi=\pi$ due to transverse momentum conservation in the hard function $\bm{\mathcal{\widetilde{H}}}^S_2$. For the hemisphere soft function in \eqref{sigbarefinal}, we set $n_1=n$ so that we only need
\begin{equation}
 g_0(0,\hat\theta,\Delta\phi) = -2\ln^2(1-\hat\theta^2) \, .
 \end{equation}
 To evaluate the color structure for the soft function with three legs explicitly, one can use the relation
 \begin{equation}
-2 \,\bm{T}_0\cdot\bm{T}_1 = \bm{T}_0^2 + \bm{T}_1^2 - \bm{T}_2^2
 \end{equation}
 which follows from color conservation $\sum_{i=0}^2  \bm{T}_i=0$ together with $\bm{T}_i^2 = C_i\, \bm{1}$, where $C_i$ is the quadratic Casimir of the relevant representation, $C_q=C_F$ and $C_g=C_A$.
 
 For $\bm{\mathcal{\widetilde{S}}}_1$ in the left-jet case, we can set $n_1=n$  ($\theta_1=0$) since the hard function will enforce that the single hard parton must fly along the thrust axis. For completeness, we reproduce the two-loop result for this function given in  \cite{Becher:2016mmh}. Using relation \eqref{eq:relCoft} we have
 \begin{equation}\label{eq:softOne}
  \langle \bm{\mathcal{\widetilde{S}}}_1(\{n\}, \tau_L,\mu) \rangle
   = 1 + \frac{C_F\alpha_s}{4\pi} \left( -4 L_L^2 - \frac{\pi^2}{2} \right)
    + \left(\frac{\alpha_s}{4\pi}\right)^2 \left( C_F^2\,u_1^F + C_F C_A\,u_1^A + C_F T_F n_f\,u_1^f \right) ,
\end{equation}
where $L_L=\ln(\tau_L/\mu)$ and
\begin{align}
   u_1^F &= 8 L_L^4 + 2\pi^2 L_L^2 + \frac{\pi^4}{8} \,, \nno \\
   u_1^A &= \frac{88 L_L^3}{9} - \frac{268 L_L^2}{9}
    + \left( \frac{844}{27} - \frac{22\pi^2}{9} - 28\zeta_3 \right) \!L_L 
    - \frac{836}{81} - \frac{1139\pi^2}{108} - \frac{187\zeta_3}{9} +\frac{4\pi^4}{5} \,, \nno \\
   u_1^f &= - \frac{32 L_L^3}{9} + \frac{80 L_L^2}{9} 
    + \left( - \frac{296}{27} + \frac{8\pi^2}{9} \right) \!L_L - \frac{374}{81} + \frac{109\pi^2}{27}
    + \frac{68\zeta_3}{9} \,.
\end{align}
The renormalization of the soft function is quite non-trivial since
higher-multiplicity function mix into lower ones, see
\eqref{eq:SmZfac}. It is therefore interesting to test that the
renormalization factor, obtained from absorbing the divergences of the
hard functions, indeed renders the soft functions finite. For the case
of narrow-jet cross sections, this was verified in
\cite{Becher:2016mmh}. Since we work with different hard functions in
the present case, it is an important but somewhat tedious exercise to
show that one recovers the same soft function after performing the
renormalization. We have checked that this is the case -- the details
can be found in Appendix \ref{sec:renorm}.

\subsection{Hard functions}
Since $\bm{\mathcal{\widetilde{H}}}^S_0(\tau_R,\epsilon )=\bm{1}$ is trivial, the first nontrivial hard function 
is  $\bm{\mathcal{\widetilde{H}}}^S_1(\{n_1\},\tau_R,\e)$,
 which arises from the emission of a single hard gluon from the Wilson-line operator in \eqref{hardAmpS}. 
The leading contribution to this hard function is given by
\begin{align}
\label{eq:H1integrand}
\frac{\alpha_s}{4\pi}\bm{\mathcal{\widetilde{H}}}_1^{S(1)}(\{n_1\},\tau_R,\e) = &\frac{2C_F g_s^2\,\tilde{\mu}^{2\e}}{(2\pi)^{2-2\e}} \int_0^{\infty } d\omega_R   \int dE_1 E_1^{1-2\e} \frac{n\cdot \bar n}{n\cdot p_1 \bar n \cdot p_1} \theta(\bar n\cdot p_1 - n\cdot p_1) \nonumber \\
& \times e^{-\omega_R/(\tau_R e^{\gamma_E})} \delta(\omega_R - n\cdot p_1) \bm{1} \,.
\end{align}
The light-cone vector $n_1$ appearing as an argument in the hard function is related to the gluon
momentum according to $p_1^\mu = E_1 n_1^\mu$. We parameterize this vector in $d$-dimensions 
as $n_1 = (1,0,\dots, \cos\theta_1)$, so that
the theta-function constraint  in (\ref{eq:H1integrand}) gives support  to the hard function only in
the region $0<\cos\theta_1<1$, that is, when the gluon is in the right hemisphere.  After integrating over 
$E_1$ and $\omega_R$ and performing the trivial angular integrations, we are left with an angular convolution in $\theta_1$. It is convenient to instead use the angular variable $\hat{\theta}_1$ defined in \eqref{hatTheta} and write
\begin{align}
 \bm{\mathcal{\widetilde{H}}}^{S(1)}_1(\{ n_1\},\tau_R,\e) \otimes \bm{\mathcal{\widetilde{S}}}_2^{(1)}(\{ n_1\},\tau_L,\e) &= \int \frac{d\Omega(n_1)}{4\pi} 
  \bm{\mathcal{\widetilde{H}}}^{S(1)}_1(\{ n_1\},\tau_R,\e) \, \bm{\mathcal{\widetilde{S}}}_2^{(1)}(\{ n_1\},\tau_L,\e)\nno \\
  & = 
\int_0^1 d\hat{\theta}_1 \bm{\mathcal{\widetilde{H}}}_1^{S(1)}(\hat{\theta}_1,\tau_R,\e) \bm{\mathcal{\widetilde{S}}}_2^{(1)}(\hat{\theta}_1,\tau_L,\e)\,,
\end{align}
where we have absorbed the trivial part of the angular integration into $ \bm{\mathcal{\widetilde{H}}}_1^{S(1)}(\hat{\theta}_1,\tau_R,\e)$.  For the bare hard function at NLO, we obtain the simple result
\begin{equation}
\bm{\mathcal{\widetilde{H}}}_1^{S(1)}(\hat{\theta}_1,\tau_R,\e) =   8 C_F\, \left(\frac{\mu}{\tau_R}\right)^{2\epsilon} \,
\frac{e^{-\epsilon \gamma_E}\Gamma(-2\epsilon)}{\Gamma(1-\epsilon)} \, \hat{\theta}_1^{-1+2\epsilon} \,\bm{1}   \,.
\end{equation}
The hard function is thus a distribution in the angle $\hat{\theta}_1$, in contrast the soft function which is regular for $\hat{\theta}_1\to 0$ . To obtain the renormalized hard function, one uses the identity
\begin{equation}\label{eq:distexp}
\hat{\theta}_1^{-1+2\epsilon}= \frac{1}{2\epsilon} \delta(\hat{\theta}_1) + \left[ \frac1{\hat{\theta}_1} \right]_+  + 2\epsilon \left[ \frac{\ln \hat{\theta}_1}{\hat{\theta}_1} \right]_+ + \dots \, .
\end{equation}
The renormalized one-loop function is given by
 \begin{align}
\bm{\mathcal{\widetilde{H}}}_1^{S(1)}(\hat{\theta}_1,\tau_R,\mu) &= \bm{\mathcal{\widetilde{H}}}_1^{S(1)}(\hat{\theta}_1,\tau_R,\e) - \bm{\mathcal{\widetilde{H}}}_0^{S(0)}(\tau_R,\e) \bm{\mathcal{ \widetilde{Z}}}^{(1)}_{01}(\hat{\theta}_1 ,\tau_R, \e,\mu)\\
& =\bm{\mathcal{\widetilde{H}}}_1^{S(1)}(\hat{\theta}_1,\tau_R,\e) - \bm{\mathcal{ \widetilde{Z}}}^{(1)}_{01}(\hat{\theta}_1 ,\tau_R, \e,\mu)\,.
\end{align}
At this order, renormalization is equivalent to dropping the divergences in the bare function.
Doing so leaves the finite result
\begin{align}\label{H1ren}
\bm{\mathcal{\widetilde{H}}}_1^{S(1)}( \hat{\theta}_1,\tau_R,\mu)=& C_F \left\{\left(-4 L_R^2 - \frac{\pi^2}{2}\right) \delta(\hat{\theta}_1) + 8  L_R \left[ \frac1{\hat{\theta}_1} \right]_+ - 8 \left[ \frac{\ln \hat{\theta}_1}{\hat{\theta}_1} \right]_+  \right\}
\,\bm{1}  \, ,\nonumber \\
\end{align}
with $L_R= \ln(\tau_R/\mu)$. 

Finally, we also need $\bm{\mathcal{\widetilde{H}}}^{S(2)}_1$, the
one-loop correction to the one-emission function, as well as the
leading-order two-emission function
$\bm{\mathcal{\widetilde{H}}}^{S(2)}_2$. Both of these are
$\mathcal{O}(\alpha_s^2)$ corrections. Rather than computing the full
functions, it is sufficient to obtain the angular convolution of these
functions with the trivial leading-order soft functions. The bare
results for these can be extracted from the computations in
\cite{Kelley:2011ng, Hornig:2011iu} and are given in Appendix
\ref{sec:Barefun}. After renormalization one obtains 
\begin{align}\label{eq:hard2}
\bigg\langle \Big[ &\Htilde{1}^{S(2)}(\{\underline{n}\},\tau_R,\mu)+ 
\Htilde{2}^{S(2)}(\{\underline{n}\},\tau_R,\mu)\Big] \otimes \bm {1}  \bigg\rangle
 = C_F^2 \left[8 L_R^4 +2\pi^2 L_R^2 +\frac{\pi^4}{8} \right]
\nonumber \\ 
& +C_A C_F \bigg[\frac{88}{9}L_R^3-\frac{268}{9}L_R^2
+\left(\frac{772}{27}+\frac{22\pi^2}{3}-20\zeta_3  \right)L_R
 -\frac{1196}{81}-\frac{67\pi^2}{12} +\frac{17\pi^4}{45}
-\frac{319\zeta_3}{9}\bigg] 
 \nonumber \\
& +C_F T_F n_f\left[ -\frac{32}{9}L_R^3+\frac{80}{9}L_R^2
-\left(\frac{152}{27}+\frac{8\pi^2}{3}\right) L_R+\frac{238}{81}+\frac{5\pi^2}{3}+\frac{116\zeta_3}{9}  \right] ,
 \end{align}
 as is shown in Appendix \ref{sec:renorm}.   
 
\subsection{Renormalized results to NNLO}

Using \eqref{eq:softOne} and  \eqref{H1ren}, we immediately obtain the renormalized hemisphere soft function at NLO, which is given by
\begin{align}
\tilde{s}^{(1)}(\tau_L,\tau_R, \mu) &= \big\langle \bm{\mathcal{\widetilde{S}}}_1^{(1)} \big\rangle  + \big\langle \bm{\mathcal{\widetilde{H}}}^{S(1)}_1 \otimes \bm{1} \rangle = C_F \left( -4 L_L^2 -4 L_R^2 - \pi^2 \right) .
\end{align}
We observe that after the substitution $\tau_R \to \tau_L$, the hard function contribution, given by the coefficient of
the delta-function term in (\ref{H1ren}), agrees with the soft
function contribution given in \eqref{eq:softOne}. This is easily
understood since both arise from the same Wilson line matrix element
and the single emission is always left for the soft function and right
in the case of the hard function. This simple symmetry is no longer 
present at the two-loop level, since soft gluons can radiate to the
right, while hard partons cannot enter the left hemisphere.

To obtain the NNLO result, we also need the convolution of $\bm{\mathcal{\widetilde{H}}}^S_1$ with the one-loop soft function. It is easy to show that
\begin{align}\label{eq:mixed}
 \left\langle \Htilde{1}^{S(1)}(\hat{\theta}_1,\tau_R,\mu)\otimes \Stilde{2}^{(1)}(\hat{\theta}_1,\tau_L,\mu) \right\rangle  =  & \,  C_F^2 \left[ 16L_L^2L_R^2 + 2\pi^2 L_L^2 + 2\pi^2 L_R^2 + \frac{\pi^4}{4}\right]  \nno \\ &
 \hspace{-2cm} + C_A C_F \left[\frac{8\pi^2}{3} L_L L_R+8\zeta_3(L_L-2 L_R) -\frac{\pi^4}{45} \right] .
\end{align}
With the final ingredient in place, we can now evaluate \eqref{eq:NNLOfact} by adding \eqref{eq:softOne}, \eqref{eq:hard2} and \eqref{eq:mixed}. Explicitly, we have 
\begin{align}
\tilde{s}^{(2)}(\tau_L,\tau_R,\mu) =&\,  C_F^2 \frac{1}{2}\left[  4 L_L^2+ 4 L_R^2+\pi ^2 \right]^2 +C_F C_A \Bigg[ \frac{88 }{9} \left(L_L^3+L_R^3\right) -\frac{268 }{9} \left(L_L^2 + L_R^2\right) \nno \\
& \hspace{-1.5cm} +\frac{8}{3} \pi ^2 L_L L_R + \left(\frac{844}{27} -\frac{22 \pi ^2 }{9} -20  \zeta_3\right) L_L   + \left(\frac{772}{27} +\frac{22 \pi ^2 }{3} - 36  \zeta_3\right) L_R \nno \\
& \hspace{-1.5cm} -\frac{2032}{81} -\frac{871 \pi ^2}{54} -\frac{506 \zeta_3}{9}+\frac{52 \pi ^4}{45} \Bigg] + C_F T_F n_f \Bigg[ -\frac{32}{9}  \left(L_L^3 + L_R^3\right) +\frac{80}{9}  \left(L_L^2+L_R^2\right) \nno \\
& \hspace{-1.5cm}   - \left( \frac{296}{27} - \frac{8\pi^2}{9} \right)L_L -\left( \frac{152}{27} + \frac{8\pi^2}{3} \right)L_R -\frac{136}{81}+\frac{154 \pi ^2}{27}
+\frac{184
   \zeta_3}{9} \Bigg].
\end{align}
This result is equivalent to a result for the integrated soft function given in \cite{Kelley:2011ng}, and to a position-space expression given in \cite{Hornig:2011iu}. In those references the full hemisphere soft function was evaluated, while we directly obtain the function in the limit $\tau_L \ll \tau_R$. The agreement provides a nontrivial check on our factorization formula \eqref{sigbarefinal}. We have performed similar two-loop checks in our earlier work on jet cross sections. However, in that case we could only compare against numerical results from fixed-order event generators. The present case has the advantage that we can compare against the analytical results from \cite{Kelley:2011ng,Hornig:2011iu}.

In earlier work on the hemisphere soft function 
\cite{Hoang:2008fs, Kelley:2011ng, Hornig:2011iu}, the result was typically written in the form
\begin{align}\label{eq:Sng}
\widetilde{s}(\tau_L,\tau_R,\mu)=\widetilde{s}_\mu(\tau_L,\mu)\widetilde{s}_\mu(\tau_R,\mu)\widetilde{s}_{\rm ng}(r) \, .
\end{align}
The non-global remainder $\widetilde{s}_{\rm ng}(r)$ is $\mu$-independent but contains logarithms of the small ratio $r=\tau_L/\tau_R \ll 1$. As it stands, the definition of the non-global piece in \eqref{eq:Sng} is not unique. One way to fully specify it is to set $\widetilde{s}_\mu(\tau,\mu) =  \sqrt{\widetilde{s}(\tau,\tau,\mu)}$.
Dividing out the global pieces from our result, we are then left with
\begin{align}
\label{eq:AnsNG}
\widetilde{s}_{\rm ng}(r) = 1+ \left(\frac{\alpha_s}{4\pi}\right)^2\left[C_F C_A s_{{\rm ng},A}+ C_F T_F n_fs_{{\rm ng},f}  \right],
\end{align}
where 
\begin{align}
 s_{{\rm ng},A}& = -\frac{4\pi^2}{3}\ln^2(r) +\left( \frac{4}{3}-\frac{44\pi^2}{9}+8\zeta_3\right)\ln(r)
  \nonumber \, , &
 s_{{\rm ng},f}& = \left(-\frac{8}{3}+\frac{16\pi^2}{9} \right)\ln(r) \,.
\end{align}
Equally well, we could have defined the global part
$\widetilde{s}_\mu(\tau,\mu) $ as the square root of the thrust soft
function or the solution of the RG equation for $\widetilde{s}_\mu(\tau,\mu)$
with trivial boundary condition $\widetilde{s}_\mu(\tau,\tau)=1$. With
the latter two definitions, the non-global piece would involve
constant terms.

The reasoning for splitting the soft function into global
and non-global parts was that the global piece follows from the RG
evolution of the soft function $\widetilde{s}(\tau_L,\tau_R,\mu)$,
while the logarithms in the non-global part do not. However, we have
completely factorized this soft function in \eqref{sigbarefinal}. Our
factorization theorem splits the function into contributions from
$\bm{\mathcal{H}}^S_m$, which live at the scale $\tau_R$, and
contributions from $\bm{\mathcal{S}}_m$,  which live at the low scale
$\tau_L$. The RG equations for these functions simultaneously resum
{\em all} logarithms in the hemisphere soft function. So from the point of
view of our effective theory, the splitting into global and non-global
logarithms is artificial. The intricate structure of the logarithms
is simply a reflection of the complicated operator structure in the
effective theory.

\section{Logarithmic corrections to the light-jet mass distribution at NNLO \label{sec:lightjet}}

We can obtain the logarithmic corrections to the light-jet mass distribution from those for the heavy-jet and left-jet mass
distributions using (\ref{eq:lightjet}).  Since the NNLO corrections to the heavy-jet distribution are known,   
we first give new results for the NNLO corrections to the left-jet mass,  before 
converting them into results for the light-jet mass and comparing with numerical results from 
event generators at the end of the section.

The factorization theorem for the left-jet mass distribution was given
in \eqref{eq:factlightmass}. It is again convenient to work in Laplace
space since the convolution with the jet function turns into an
ordinary product. Introducing the Laplace transformation as in
\eqref{eq:LaplaceS} the cross section becomes 
\begin{equation}\label{eq:factlightmassLap}
\tilde\sigma(\tau_L)  = \sum_{i=q,\bar{q},g} \tilde{j}_i(\tau_L Q) \, \sum_{m=1}^\infty \big\langle \bm{\mathcal{H}}^i_m(\{\underline{n}\}, Q) \otimes \bm{\mathcal{\widetilde{S}}}_m(\{\underline{n}\},\tau_L) \big\rangle \,.
\end{equation}
The Laplace-transformed jet functions $\tilde{j}_i$ are the standard inclusive jet functions, which are well known. The soft functions are the same as the ones for the hemisphere soft case and were given in Section~\ref{sec:soft}. This leaves us with a computation of the relevant hard functions and the evaluation of the angular integrals over the directions of the reference vectors.

The definition of the hard functions $\bm{\mathcal{H}}_m^{i}$ for the left-jet mass, given in (\ref{eq:LightJetHard}),  involves matrix elements with a single hard parton of flavor $i=q,\bar{q},g$ on the left and $m$ hard partons on the right. The $m=1$ hard functions have the form
\begin{equation}\label{eq:hardOne}
 \bm{\mathcal{H}}_1^{q}(\hat{\theta}_1,Q,\mu)  =  \bm{\mathcal{H}}_1^{\bar{q}}(\hat{\theta}_1,Q,\mu) = \frac{\sigma_0}{2} \delta(\hat{\theta}_1) H(Q^2,\mu)  \bm{1}\,,
 \end{equation}
 where $\sigma_0$ is the Born cross section for $\gamma^*\to q\bar{q}$ decay, given in $d$-dimensions by
 \begin{align}
 \sigma_0= 3 \,\alpha \, Q_f^2 \, Q \frac{ (4\pi)^\e \,\Gamma(2-\e)}{\Gamma(2-2\e)}\left(\frac{\mu}{Q}\right)^{2\e},
 \end{align}
with $\alpha=e^2/(4\pi)$ the fine structure constant and $Q_f$ the
charge of the quark flavor $q$.  Moreover, $H(Q^2,\mu)$ is the
standard dijet hard function present also in \eqref{eq:factMLMR}, and
the $\delta$-function in the angle arises because momentum
conservation enforces that $n_1=n$. The factor $1/2$ is present
because it is arbitrary whether we label the quark or anti-quark as
being in the left hemisphere, so the two situations are averaged over.  

We also need the hard functions for the case of two hard partons in the right hemisphere. For the case of a quark-jet in the left hemisphere, we have
\begin{align}\label{eq:H2q}
\bm{\mathcal{H}}_2^{q(1)}(&\hat{\theta}_1,\hat{\theta}_2,\Delta\phi,Q,\e) =  2 \sigma_0 C_F \left(\frac{\mu}{Q}\right)^{2\epsilon} \,
\frac{e^{\epsilon \gamma_E}}{\Gamma(1-\epsilon)}
 \hat{\theta }_1^{-1-2 \epsilon } \hat{\theta }_2^{-2
   \epsilon } \left(\hat{\theta }_1+\hat{\theta }_2\right)^{-3+2 \epsilon }
   \left(1-\hat{\theta }_1 \hat{\theta }_2\right)^{-2 \epsilon } \nonumber \\
   & \times \left[2 \hat{\theta}_2 \left(1-\hat{\theta}_1^2\right) \left( 1 - \hat{\theta}_1\hat{\theta}_2\right) \left(\hat{\theta}_1 + \hat{\theta}_2 \right)+  (1-\epsilon)\hat{\theta }_1^2 \left(1+\hat{\theta }_2^2\right)^2 \right] \delta(\Delta \phi-\pi)  \bm{1}\,,
\end{align}
where $\hat{\theta }_1$ is the anti-quark angle and $\hat{\theta }_2$ the one of the gluon. Momentum conservation enforces $\Delta \phi=\phi_2-\phi_1=\pi$, which is why we only computed the soft function for this configuration. The thrust-axis constraint imposes the conditions 
\begin{align}\label{eq:angleConstr}
\sqrt{1+ \hat{\theta}_1^2} &> \hat{\theta}_1+ \hat{\theta}_2\, , & \sqrt{1+ \hat{\theta}_2^2} &> \hat{\theta}_1+ \hat{\theta}_2\,,
\end{align}
on the angular integration region, which can be added as $\theta$ functions to \eqref{eq:H2q}. This constraint implies in particular that the smaller of the two angles $\hat{\theta}_1$ and $\hat{\theta}_2$ must be less than $1/\sqrt{3}$, which corresponds to a $60^\circ$ angle from the thrust axis. When the limit is reached the three partons are in a symmetric configuration and have all the same energy. If the angle becomes larger the thrust axis flips, since it always points in the direction of the most energetic parton in a three-parton configuration. For $\epsilon \to 0$, the function $\bm{\mathcal{H}}_2^{q}$ has overlapping divergences
when  the angles $\hat{\theta }_1$ and $\hat{\theta }_2$ go to zero simultaneously. To treat these, one splits the angular integration  into two sectors $\hat{\theta }_1< \hat{\theta }_2$ and $\hat{\theta }_1> \hat{\theta }_2$ and then parametrizes $\hat{\theta }_1=u\, \hat{\theta }_2$ with $u=0\dots 1$ in the first sector and conversely in the second one. Once the divergences are separated one can expand both functions in $\epsilon$ using the identity \eqref{eq:distexp} in the appropriate variables. At the one-loop level the renormalized expressions can be obtained by simply dropping the divergences which arise in this expansion.

The second configuration which is relevant is the one where we have a gluon jet on the left and a hard $q\bar{q}$ pair on the right. The hard function for this case reads
\begin{align}
\bm{\mathcal{H}}_2^{g(1)}(&\hat{\theta}_1,\hat{\theta}_2,\Delta\phi,Q,\e) = 2 \sigma_0 C_F \left(\frac{\mu}{Q}\right)^{2\epsilon} \, \frac{e^{\epsilon \gamma_E}}{\Gamma(1-\epsilon)}
\hat{\theta }_1^{-2 \epsilon } \hat{\theta }_2^{-2 \epsilon } \left(\hat{\theta}_1+\hat{\theta }_2\right)^{-2+2 \epsilon } \left(1-\hat{\theta }_1 \hat{\theta}_2\right)^{-1-2 \epsilon}\nonumber\\
& \left[\left(1+\hat{\theta }_2^4\right) \hat{\theta }_1^2+\left(1+\hat{\theta}_1^4\right) \hat{\theta }_2^2+4 \hat{\theta }_2^2 \hat{\theta}_1^2 -  \epsilon \left( \hat{\theta}_1+\hat{\theta }_2 \right)^2  \left(1-\hat{\theta }_1 \hat{\theta}_2\right)^{2} \right]\, \delta(\Delta \phi-\pi) \bm{1}
\end{align}
and is subject to the same angular constraints \eqref{eq:angleConstr}. This hard function does not suffer from divergences 
when the angles go to zero, so we can immediately set $\epsilon \to 0$.

To obtain the full NNLO result for the left-hemisphere cross section, we would need also the one-loop corrections to $\bm{\mathcal{H}}_2^{i}\otimes \bm{1}$ and the three-parton functions $\bm{\mathcal{H}}_3^{i}\otimes \bm{1}$. However, if we are only interested in the logarithmic terms, we can avoid their computation by setting $\mu=Q$. For this scale choice these functions do not contain any logarithms and we can therefore recover the logarithmic part of the NNLO cross section from 
\begin{align}\label{eq:sigLap}
 \tilde\sigma(\tau_L) = & \, 2 \,\tilde{j}_q(\tau_L Q,\mu)  \left\langle \bm{\mathcal{H}}_1^{q}(\{n_1\},Q,\mu)   \otimes \bm{\mathcal{\widetilde{S}}}_1(\{n_1\}, \tau_L,\mu) \right\rangle  \nonumber \\
 &+ \sum_{i=q, \bar{q}, g} \tilde{j}_i(\tau_L Q,\mu)  \left\langle \bm{\mathcal{H}}_2^{i}(\{n_1,n_2\} ,Q,\mu)   \otimes \bm{\mathcal{\widetilde{S}}}_2(\{n_1,n_2\}, \tau_L,\mu) \right\rangle\, + \mathcal{O}(\alpha_s^2 L_L^0)\,,
\end{align}
where the factor 2 in the first line accounts for the identical
contribution when the anti-quark is in the left hemisphere. The
two-loop result for the soft function $\bm{\mathcal{\widetilde{S}}}_1$
was given in the previous section in \eqref{eq:softOne}. The dijet
hard function \eqref{eq:hardOne} and the Laplace-space quark jet
function $\tilde{j}_q$ are well known. Explicit two-loop results for
both quantities can be found in Appendix B of \cite{Becher:2008cf}. We
can thus immediately evaluate the first line of \eqref{eq:sigLap} and
what remains is the convolution on the second line. Since the
functions $\bm{\mathcal{H}}_2^i$ start at
$\mathcal{O}(\alpha_s)$, we need the gluon jet function $\tilde{j}_g$ and the soft function $\bm{\mathcal{\widetilde{S}}}_2$
only to  one-loop order.

We have obtained analytical results for the convolutions of the two-parton functions with the trivial leading-order soft functions
\begin{align}
&\sum_{i=q,\bar{q}}\left\langle \bm{\mathcal{H}}_2^{i(1)}(\{n_1,n_2\} ,Q,\mu)   \otimes \bm{1} \right\rangle = C_F \sigma_0  \Bigg[ 4\,L_Q^2 - 6\,L_Q + \frac{29}{3} - \frac{3\,\pi^2}{2} - 2\ln^2 2 \nonumber\\
& \hspace{9.45cm}+ \frac{5}{4}\ln\,3 - 4\,\rm{Li}_2\left(-\frac{1}{2}\right) \Bigg]  \, , \label{eq:Hq}\\
&\left\langle \bm{\mathcal{H}}_2^{g(1)}(\{n_1,n_2\} ,Q,\mu)   \otimes \bm{1} \right\rangle  = C_F \sigma_0 \Bigg[ -\frac{1}{6}+\frac{\pi ^2}{3}+2 \ln ^2 2 -\frac{5 }{4} \ln 3 + 4\, \text{Li}_2\left(-\frac{1}{2}\right) \Bigg] \, , \label{eq:Hg}
\end{align}
where $L_Q = \ln\left(Q/\mu \right)$. The appearance of logarithms and polylogarithms in addition to the usual $\zeta$-values is a result of the phase-space constraint \eqref{eq:angleConstr}. The result in \eqref{eq:Hg} agrees with the quantity $r_3$ obtained in \cite{Burby:2001uz}, see (22) in \cite{Dasgupta:2001sh}. Putting \eqref{eq:Hq} together with the other one-loop ingredients we obtain agreement with the result of \cite{Dasgupta:2001sh} also in the quark channel. For the NNLO cross section we  need results for the convolutions with the NLO soft function \eqref{eq:SmNLO}, which have the form 
\begin{align}\label{eq:renConvq}
\sum_{i=q,\bar{q}} \Big\langle \bm{\mathcal{H}}_2^{i(1)}&(\{n_1,n_2\} ,Q,\mu)   \otimes \bm{\mathcal{\widetilde{S}}}_2^{(1)}(\{n_1,n_2\}, \tau_L,\mu)  \Big \rangle = \nonumber \\ 
&C_F^2 \sigma_0\Bigg[ \left(-16 \, L_Q^2 + 24 \,  L_Q + M_{q,F}^{(2)} \right) L_L^2 + M_{q,F}^{(1)}\, L_L -2\pi^2 L_Q^2 + 3\pi^2 L_Q + M_{q,F}^{(0)}\Bigg]   \nno \\
&\hspace{1cm}+ C_F C_A \sigma_0\Bigg[ \left( \frac{8\pi^2 L_Q}{3} +M_{q,A}^{(1)}\right) L_L  - 16\zeta_3 L_Q + M_{q,A}^{(0)} \Bigg]  \, ,\\
\Big\langle \bm{\mathcal{H}}_2^{g(1)}&(\{n_1,n_2\} ,Q,\mu)   \otimes \bm{\mathcal{\widetilde{S}}}_2^{(1)}(\{n_1,n_2\}, \tau_L,\mu)  \Big\rangle  =   \nno \\
&  C_F^2 \sigma_0 \left[ M_{g,F}^{(1)} L_L+ M_{g,F}^{(0)} \right] + C_F C_A \sigma_0 \left[ M_{g,A}^{(2)} L_L^2  + M_{g,A}^{(1)} L_L   + M_{g,A}^{(0)} \right]  . \label{eq:renConvg}
\end{align}
The expressions for the coefficients $M_{g,F}^{(i)}$ and $M_{q,F}^{(i)}$ are lengthy and can be found in the appendix in \eqref{eq:McoeffsRen}.

Putting everything together and inverting the Laplace transformation we then obtain all logarithmic terms in the left-jet mass distribution.  The inverse Laplace transformation can be obtained using the simple substitution rules
\begin{align}
&\ln \frac{\tau_L}{Q} \to \ln \! \rho_L,~~~ \ln^2 \frac{\tau_L}{Q} \to \ln^2 \! \rho_L - \frac{\pi^2}{6},~~~ \ln^3 \frac{\tau_L}{Q} \to \ln^3 \! \rho_L - \frac{\pi^2}{2} \ln  \! \rho_L + 2\zeta_3, \nno \\
& \ln^4 \frac{\tau_L}{Q} \to \ln^4 \! \rho_L - \pi^2 \ln^2 \! \rho_L + 8 \zeta_3 \ln\!\rho_L + \frac{\pi^4}{60}.
\end{align}
Using relation \eqref{eq:lightjet} together with the known result for the logarithmic terms in the heavy-jet mass distribution \cite{Chien:2010kc} we then obtain the light-jet mass distribution. Up to NNLO, it has the general form
\begin{equation}
\frac{1}{\sigma_0} \frac{d\sigma}{d\rho_\ell} = \delta(\rho_l) \, \left\{1 + \left(\frac{\alpha_s}{2\pi}\right)\frac{3 C_F}{2}  + \left(\frac{\alpha_s}{2\pi}\right)^2 B_\delta \right\}+\left(\frac{\alpha_s}{2\pi}\right)^2 \left[ \frac{B_+(\rho_l)}{\rho_l}\right]_+ + \cdots
\, . 
\end{equation}
Note that at NLO, the distribution is a $\delta$-function since the lighter jet contains only a single parton. A nontrivial light-jet mass distribution first arises from four-particle configurations at NNLO in which each hemisphere contains two partons. The logarithmic terms from these configurations are encoded in the function $B_+(\rho_\ell)$, for which we obtain
\begin{align}\label{eq:Bpcoeff}
B_+(\rho) = \, & C_F^2\Bigg[ -4 \ln^3\rho - 9 \ln^2\rho + \left[ -\frac{59}{6}+\frac{4\, \pi ^2}{3}+4 \ln^2 2-\frac{5 \ln 3 }{2} + 8 \,\text{Li}_2\left(-\frac{1}{2}\right) \right] \ln\rho \nno \\
&  \hspace{.8cm} + \frac{15}{2}  +2 \pi ^2 +\frac{809 \zeta_3 }{6}  +\frac{88 \ln ^3 2}{3}+8 \ln 2 \ln
   ^2 3+\frac{5 \ln ^2 3}{2}-24 \ln ^2 2 \ln 3+\frac{27 \ln
   ^2 2}{2} \nno \\
   &  \hspace{.8cm} -28 \ln 2 \ln 3 +\frac{487 \ln 3}{24}-\frac{20}{3} \pi ^2
   \ln 2-\frac{88 \ln 2}{3} +43 \,
   \text{Li}_2\left(-\frac{1}{2}\right)  -16 \,  \text{Li}_2\left(-\frac{1}{2}\right)
   \ln 3 \nno \\
   &  \hspace{.8cm}+96 \, \text{Li}_2\left(-\frac{1}{2}\right) \ln 2 -8 \, \text{Li}_3\left(\frac{3}{4}\right)+176 \, 
   \text{Li}_3\left(-\frac{1}{2}\right) -8 \, I_2  \Bigg] \nno \\
& \hspace{-.8cm}+  C_F C_A \Bigg[ \left[ \frac{1}{3}-2 \pi ^2-4 \ln^2 2 +\frac{5
   \ln 3}{2} -8 \,\text{Li}_2\left(-\frac{1}{2}\right) \right]  \ln\rho -\frac{407}{72}  -\frac{13 \pi ^2}{18} -\frac{389 \zeta_3}{3} -\frac{8 \ln ^3 3}{3} \nno \\
   & \hspace{.8cm}  -52 \ln
   ^3 2-12 \ln 2 \ln ^2 3-\frac{15 \ln ^2 3}{4}+52 \ln ^2 2 \ln
   3+\frac{43 \ln ^2 2}{12}-\frac{11}{2} \ln 2 \ln 3 \nno \\
   & \hspace{.8cm} -\frac{917 \ln
   3}{24}+6 \pi ^2 \ln 2+\frac{212 \ln 2}{3} +20 \, \text{Li}_3\left(\frac{3}{4}\right)  +\frac{235
   }{6} \, \text{Li}_2\left(-\frac{1}{2}\right)  \nno \\
   & \hspace{.8cm}+24 \,
   \text{Li}_2\left(-\frac{1}{2}\right) \ln 3-88 \,
   \text{Li}_2\left(-\frac{1}{2}\right) \ln 2  +16 \, 
   \text{Li}_3\left(\frac{1}{3}\right)-112 \, 
   \text{Li}_3\left(-\frac{1}{2}\right)  -8 \, I_1 \Bigg] \nno \\
   & \hspace{-.8cm} + C_F T_F n_f \Bigg[ -\frac{13}{9}+\frac{10\, \pi
   ^2}{9}+\frac{4 }{3}\ln ^2 2-\frac{5 }{6}\ln3 + \frac{8 }{3} \, \text{Li}_2\left(-\frac{1}{2}\right) \Bigg]\,.
\end{align}
Due to the uncalculated two-loop constant terms in the hard functions
$\bm{\mathcal{H}}_2$ and $\bm{\mathcal{H}}_3$,
we cannot give the two-loop coefficient $B_\delta$, but the
$\delta$-function terms do not contribute to the logarithmic
corrections to the light-jet mass distribution. 
We have verified that the terms involving powers of $\ln \rho$ in (\ref{eq:Bpcoeff})
are in agreement with those implied by the results of \cite{Burby:2001uz,Dasgupta:2001sh}.  
The remaining pieces,  on the other hand, are new.  As a further check,  we have
repeated the computation of the logarithmic terms in the cross section
using bare instead of renormalized quantities. The logarithms are
related to divergences in the individual ingredients in the
factorization theorem \eqref{eq:factlightmass}. To obtain the
logarithmic terms in the cross section we thus insert the divergent
bare ingredients together with their associated logarithmic terms into
the Laplace-transformed version of \eqref{eq:factlightmass}. The
divergences cancel and we are left with a logarithmic structure
which agrees with \eqref{eq:Bpcoeff}. The details of this computation
can be found in Appendix \ref{sec:bareLjet}.

\begin{figure}[t!]
\centering
\hspace{-0.5cm}
\includegraphics[width=1\textwidth]{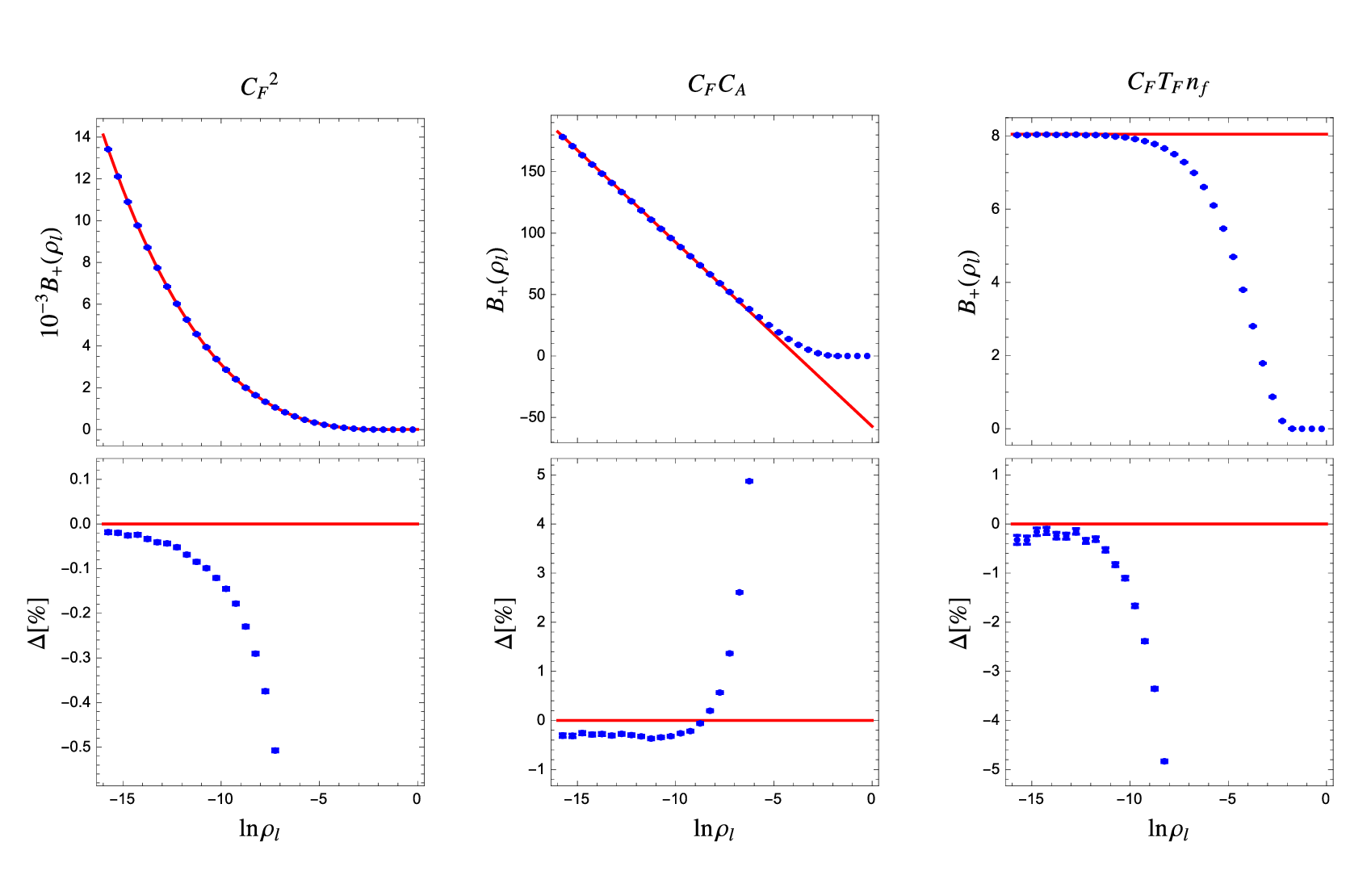}  
\vspace{-0.5cm}
\caption{Comparison of our analytic results (solid lines) for the coefficients of the three color structures in the two-loop coefficient $B_+(\rho_l)$ for the light-jet mass distribution with numerical results (points with invisibly small error bars) obtained using the {\sc Event2} event generator \cite{Catani:1996vz}. The two results must agree for small $\rho_\ell$. The lower panel shows the relative difference in per cent. \label{fig:EV2Comparison}}
\end{figure}

In contrast to the hemisphere soft function, the full analytical result for the light-jet mass distribution is not known, but our result for the coefficient $B_+(\rho_l)$ can be compared to numerical results obtained from running a fixed-order event generator. Since our results are the leading term in the limit $\rho_\ell \to 0$, we need to run the fixed-order code for very small values of $\rho_\ell$ to suppress higher-power contributions, which makes the numerics delicate. For our comparison, we  use {\sc Event2} \cite{Catani:1996vz}, which is well suited to study the region of small $\rho_\ell$ since the phase-space generation can be tuned to focus on this region. We note that the fixed-order result is known even one order higher \cite{GehrmannDeRidder:2007hr,Weinzierl:2009ms,DelDuca:2016ily} and available in the form of a public code {\sc eerad3} \cite{Ridder:2014wza}. In order to ensure that the power-suppressed terms are small, we run down to values of $\ln \rho_\ell = -16$. To ensure numerical stability, {\sc Event2} imposes a cutoff on the invariant mass of parton pairs, and we run the code in quadruple precision to be able to lower the cutoff enough to avoid cutoff effects. Figure~\ref{fig:EV2Comparison} shows the {\sc Event2} result in blue, compared to our analytic result shown as red lines. The statistical error bars on the {\sc Event2} results are barely visible, since we have generated 300 billion events. The upper panels show that the numerical results indeed approach the leading-power analytic results as the value of $\rho_\ell$ is lowered. In the lower panel, we show the difference between {\sc Event2} and the analytic result in per cent, and the two agree to better than half a per cent for low values of $\rho_\ell$. However, our statistical uncertainties are even smaller than this and we find residual deviations in all color channels which are larger than the uncertainties.  As a cross check, we have performed the same comparison against the well-known analytical result for the heavy-jet mass \cite{Chien:2010kc} and find deviations of similar size. Indeed, earlier papers have identified similar numerical issues in several variables \cite{Chien:2010kc,Becher:2015gsa,Frye:2016aiz}, so we believe that the remaining deviations are not indicative of a problem in our analytic computation. We have also compared with the results from {\sc eerad3}  and from the CoLoRFulNNLO framework \cite{DelDuca:2016ily} but were not able to achieve small enough statistical uncertainties to resolve the difference between {\sc Event2} and the analytic result.

\section{NLL resummation\label{sec:NLL}}

Our focus has been on the factorization properties of the hemisphere
soft function and the light-jet mass distribution. The factorization
theorems we derived are important because they enable the resummation
of the large logarithms. In our framework, this resummation is
achieved by solving the RG evolution equations for the ingredients of
the factorization theorem and evolving them to a common reference
scale. To perform NLL resummation, which resums the leading non-global
logarithms, one needs to evaluate the hard, jet and soft functions at
tree level and evolve them using one-loop regular anomalous
dimensions, together with the two-loop cusp anomalous
dimension. The global part of the light-jet mass distribution at NLL was
presented in \cite{Burby:2001uz} and the non-global part in the large-$N_c$ limit was computed in
\cite{Dasgupta:2001sh}, but as far as we are aware a numerical result
for the NLL resummed single-hemisphere mass distribution including NGLs was never
presented in the literature.  

The simplest way to obtain the NLL result for the left-jet mass
distribution is to choose the factorization scale as $\mu = \mu_h \sim
Q$. With this choice, the hard functions do not suffer from large
logarithms and at NLL the factorization theorem
\eqref{eq:factlightmass} simplifies to
\begin{equation}\label{eq:factlightmassSimp} 
\frac{d \sigma}{d M_L^2}  = \sigma_0 \int_0^\infty d\omega_L\,  J_q(M_L^2 - Q\,\omega_L,\mu_h) \,\big\langle  \bm{\mathcal{S}}_1(\{n\},\omega_L,\mu_h) \big\rangle \,.
\end{equation}
We have used that all higher-order hard functions are suppressed by powers
of $\alpha_s(\mu_h)$ and can be neglected at NLL. To obtain the cross section
we thus need two ingredients: the resummed quark jet function and the
soft function $\bm{\mathcal{S}}_1(\{n\},\omega,\mu_h)$ evolved to the hard
scale $\mu_h$. This soft function is the same as the NLL resummed
result for the hemisphere soft function. Indeed, choosing $\mu =
\mu_h$ and integrating $\omega_R$ up to a large value $Q\sim \mu_h$
the factorization theorem \eqref{sigbarefinal} for this quantity
at NLL accuracy reduces to
\begin{equation}
\int_0^{Q} \!d\omega_R\, S(\omega_L, \omega_R,\mu_h)  =   \big\langle \bm{\mathcal{S}}_{1}(\{n\},\omega_L,\mu_h) \big\rangle \,.
\end{equation}
This fact is of course well known and it is for this reason that the non-global
logarithms in the light-jet mass are usually studied using the
hemisphere soft function. Beyond NLL this simple relationship is no
longer valid, because the left-jet mass receives contributions from
hard radiation in the right hemisphere.

Before analyzing the soft function further, let us quote the resummed
result for the jet function at NLL. Using the Laplace-space technique
of \cite{Becher:2006nr}, one obtains
\begin{equation}\label{Jfun}
   J_q(p^2,\mu_h)
   = \exp\left[ - 4S(\mu_j,\mu_h) + 2 A_{\gamma^J}(\mu_j,\mu_h) \right] \,
    \frac{e^{-\gamma_E\eta_J}}{\Gamma(\eta_J)}\,\frac{1}{p^2}
    \left( \frac{p^2}{\mu_j^2} \right)^{\eta_J} \,,
\end{equation}
where $\eta_J=2A_\Gamma(\mu_j,\mu_h)$. Explicitly, the Sudakov
exponent $S(\mu_j,\mu_h) $ and the single logarithmic function
$A_\Gamma(\mu_j,\mu_h)$ are
\begin{equation}\label{RGEsols}
\begin{aligned}
   S(\mu_j,\mu) 
   &= \frac{\Gamma_0}{4\beta_0^2}\,\Bigg\{
    \frac{4\pi}{\alpha_s(\mu_j)} \left( 1 - \frac{1}{r} - \ln r \right)
    + \left( \frac{\Gamma_1}{\Gamma_0} - \frac{\beta_1}{\beta_0}
    \right) (1-r+\ln r) + \frac{\beta_1}{2\beta_0} \ln^2 r  \Bigg\} \,, \\
   A_\Gamma(\mu_j,\mu) 
   &   = \frac{\Gamma_0}{2\beta_0} \,\ln r \, , 
\end{aligned}
\end{equation}
where $r=\alpha_s(\mu)/\alpha_s(\mu_j)$. The result for $A_{\gamma^J}$
is obtained by replacing $\Gamma_0 \to \gamma_0^J$ in
$A_\Gamma(\mu_j,\mu) $. The relevant expansion coefficients of the
anomalous dimensions and the $\beta$-function can be found at the end of Appendix \ref{sec:Barefun}.

The resummed soft function $ \big\langle
\bm{\mathcal{S}}_{1}(\{n\},\omega_L,\mu_h) \big\rangle$ can be
obtained by solving the RG equation for the soft functions, which in Laplace space takes
the form
\begin{equation}\label{eq:softRG}
\frac{d}{d\ln\mu}\,\bm{\widetilde{\mathcal{S}}}_l(\{\underline{n} \},\tau,\mu)  =\sum_{m= l}^\infty \bm{\Gamma}^S_{lm}(\{\underline{n}\}, \tau,\mu)\, \hat{\otimes}\, \bm{\widetilde{\mathcal{S}}}_m(\{\underline{n} \},\tau,\mu) \, .
\end{equation}
Due to the factorization theorem \eqref{eq:factlightmassLap}, the anomalous dimension matrix must take the form
\begin{equation}
\bm{\Gamma}^S_{lm}(\{\underline{n}\}, \tau,\mu) = 2 \, \Gamma_{\rm cusp} \ln\! \left(\frac{\tau}{\mu} \right) \delta_{lm}  + \bm{\hat{\Gamma}}_{lm}(\{\underline{n}\})
 \, . \label{eq:sftGamma}
\end{equation}
The cusp piece is diagonal since the $\tau$ dependence of the
anomalous dimension $\bm{\Gamma}^S_{lm}$ must cancel against that
of the jet function $\tilde{j}_q$ in \eqref{eq:factlightmassLap}. We
can thus split the soft functions into a product
\begin{equation}
\label{eq:SoftSplit}
\bm{\mathcal{\widetilde{S}}}_l(\{\underline{n} \},\tau,\mu) =  \tilde{S}_G(\tau,\mu)\, \bm{\mathcal{\hat{S}}}_l(\{\underline{n} \},\tau,\mu)\,,
\end{equation}
where the global function fulfills the simple RG equation for the cusp
part with trivial initial condition $\tilde{S}_G(\tau,\tau)=1$. In Laplace space this RG equation has the same form as for the jet function
and is easily solved.  Inverting the Laplace transformation, 
we obtain
\begin{equation}\label{SGfun}
   S_G(\omega,  \mu_h)
   = \exp\left[2S(\mu_s,\mu_h) \right] \,
    \frac{e^{-\gamma_E\eta_S}}{\Gamma(\eta_S)}\,\frac{1}{\omega}
    \left( \frac{\omega}{\mu_s} \right)^{\eta_S} \,,
\end{equation}
where $\eta_S=2 A_\Gamma(\mu_h,\mu_s)$. The remaining piece
$\bm{\mathcal{\hat{S}}}_l(\{\underline{n} \},\tau,\mu)$ in
(\ref{eq:SoftSplit}) has a single logarithmic evolution driven by
$\bm{\hat{\Gamma}}_{lm}(\{\underline{n}\})$, which can be derived from results given
in Appendix C of \cite{Becher:2016mmh}. This piece captures
the non-global logarithms, through the formal solution 
\begin{align}
\langle \bm{\mathcal{\hat{S}}}_1(\{\underline{n} \},\tau,\mu_h) \rangle &= 
\sum_{m=1}^\infty  \langle  \bm{U}_{1m}^S(\{\underline{n}\},\mu_s,\mu_h)\, \hat{\otimes}\, \bm{\mathcal{\hat{S}}}_m(\{\underline{n} \},\tau,\mu_s)\rangle &\nonumber\\
  &= \sum_{m=1}^\infty\langle \bm{U}_{1m}^S(\{\underline{n}\},\mu_s,\mu_h)\, \hat{\otimes} \, \bm{1}\rangle \equiv S_{NG}(\mu_s,\mu_h)\,,
  \end{align}
where in the second line we used  $\bm{\mathcal{\hat{S}}}_m(\{\underline{n}
  \},\tau,\mu_s)=\bm{1}+\mathcal{O}(\alpha_s)$, and made explicit 
that at NLL the quantity $S_{NG}(\mu_s,\mu_h)$ is thus a function of $\mu_h$ and
$\mu_s$ only. The evolution matrix $\bm{U}_{1m}^S$ evolves the soft function
from the low scale $\mu_s$ to the high scale $\mu_h$. 
It is obtained at NLL by exponentiating
the one-loop anomalous dimension matrix
\begin{equation}\label{eq:US}
   \bm{U}^S(\{\underline{n}\},\mu_s,\mu_h) 
   = {\bf P} \exp\left[\, \int_{\mu_s}^{\mu_h} \frac{d\mu}{\mu}\,
    \bm{\hat{\Gamma}}(\{\underline{n}\},\mu) \right] ,
\end{equation}
but due to the angular convolutions and the color structure of the
anomalous dimension matrix, deriving an explicit form for the
evolution matrix is highly nontrivial. In our paper
\cite{Becher:2016mmh} we demonstrated that in the large-$N_c$ limit
the exponentiation of the one-loop anomalous dimension matrix is
equivalent to solving the BMS equation. The RG evolution equation
\eqref{eq:softRG} is also equivalent to a parton-shower equation and
this is the way the resummation of the hemisphere soft function was
performed in the original paper of Dasgupta and Salam
\cite{Dasgupta:2001sh}, who presented a simple, accurate
parameterization of their result. In the future, it will be very
interesting to generalize this to higher logarithmic accuracy but for
the moment we will simply use their result to obtain a resummed result
for the left-jet mass and investigate the size of the leading
non-global logarithms in this observable. The parameterization of
Dasgupta and Salam has the form
\begin{equation}\label{SNG}
S_{\rm NG}(\mu_s,\mu_h)
   \approx \exp\!\left(-C_A C_F \frac{\pi^2}{3} \,u^2 \frac{1+(a u)^2}{1+(b u)^c}\right) ,  
\end{equation}
with 
\begin{equation}
u = \frac{1}{\beta_0}\ln \frac{\alpha_s(\mu_s)}{\alpha_s(\mu_h)}\,,
\end{equation}
where the constants $a=0.85\, C_A$,  $b=0.86\, C_A$, and $c=1.33$ were determined by fitting to the parton-shower result.

\begin{figure}[t!]
\begin{center}
\begin{tabular}{ccc}
\includegraphics[height=0.4\textwidth]{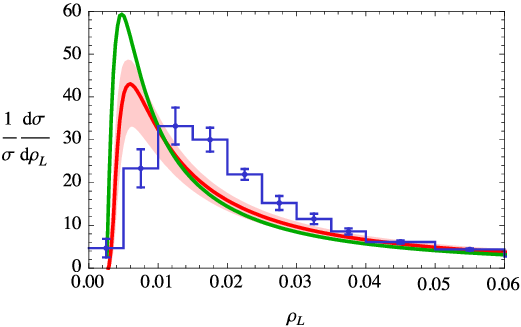} 
\end{tabular}
\end{center}
\vspace{-0.3cm}
\caption{NLL result for the left-jet mass distribution (red curve). The red uncertainty band is obtained from scale variations as explained in the text. The green line is the purely global part of the distribution. In blue we show experimental results from {\sc Aleph}  \cite{Buskulic:1992hq}.
\label{fig:NLL}}
\end{figure}

The resummed result for the soft function in momentum space is then simply the product of the global function with the non-global evolution factor,
\begin{equation}
\langle \bm{\mathcal{S}}_1(\{\underline{n} \},\omega,\mu_h) \rangle = S_{\rm NG}(\mu_s,\mu_h) S_G(\omega,\mu_h)
\, ,
\end{equation}
and the final result for the left-jet mass is obtained by convolving the soft function and the jet function. Let us first combine the global piece with the jet function. Integrating also over $\rho_L$, we obtain
\begin{align}\label{eq:Sigmaq}
\Sigma_q(\rho_L) &= \int_0^{\rho_L} d\rho_L' \int_0^{Q \rho_L'} d\omega \, J_q(Q^2 \rho_L' - Q \omega,\mu_h) S_G(\omega,\mu_h) \nonumber\\
& = \exp\left[2S(\mu_s,\mu_h)- 4S(\mu_j,\mu_h) + 2 A_{\gamma^J}(\mu_j,\mu_h) \right]\, \frac{e^{-\gamma_E  \eta }
   }{\Gamma (\eta +1)} \,\left(\frac{Q^2 \rho _L}{\mu_j^2}\right)^{\eta } \left(\frac{Q \mu _s}{\mu_j^2}\right)^{-\eta_S}\, ,
\end{align}
where $\eta=\eta_J+\eta_S=2 A_\Gamma(\mu_j,\mu_s)$. 
The integrated left-jet distribution is then obtained as
\begin{equation}\label{eq:cumulL}
R(\rho_L) = \int_0^{\rho_L} d\rho_L' \frac{1}{\sigma} \frac{d\sigma}{d\rho_L'} = S_{\rm NG}(\mu_s,\mu_h) \Sigma_q(\rho_L)\,,
\end{equation}
where we need to choose $\mu_s\sim \rho_L Q$ and $\mu_h\sim Q$. The
quantity $\Sigma_q$ plays an important role in the coherent
branching formalism \cite{Catani:1989ne,Catani:1990rr,Catani:1992ua},
where it arises as an integral over the jet function. We verified that
\eqref{eq:Sigmaq} indeed reproduces the result for this quantity given
in \cite{Burby:2001uz} after setting the scales to the default values
$\mu_j^2= \rho_L Q^2$ and $\mu_s= \rho_L Q$. Formula \eqref{eq:Sigmaq}
shows that the jet function in the coherent-branching formalism also
includes the global part of the soft radiation. Our final resummed
result \eqref{eq:cumulL} is therefore fully equivalent to that presented in
\cite{Dasgupta:2001sh}. Squaring $\Sigma_q$, one obtains the
integrated heavy-jet mass at NLL:
\begin{equation}\label{eq:cumul}
R(\rho_h) = [\Sigma_q(\rho_h)]^2\,.
\end{equation}
We have checked that using \eqref{eq:Sigmaq} in the above result reproduces the
resummed result of \cite{Chien:2010kc}. Below we will use the result
for $R(\rho_h) $ together with relation \eqref{eq:lightjet} to obtain
the light-jet mass from the left-jet mass distribution
\eqref{eq:cumulL}.

The result for the resummed left-jet mass distribution \eqref{eq:factlightmassSimp}
is shown in Figure \ref{fig:NLL}. For our plots, we choose $Q=M_Z$ and
$\alpha_s(M_Z)=0.1181$ \cite{Agashe:2014kda}. The red line shows the
result for the default scale choices, and to estimate its uncertainty,
we perform two different scale variations. In particular, we separately vary the
  hard scale $\mu_h$ and the jet scale $\mu_j$ by  factors of two
  around the default choices $\mu_h^2=Q^2$ and $\mu_j^2=\rho_L Q^2$,
  and show in the plots the envelope of the two variations. At very
  low values of $\rho_L$ the spectrum ends because $\mu_s=\rho_L Q$
  hits the Landau pole. One could also vary the soft scale, which
  would shift this end-point and thus generate a larger uncertainty band. The green line in the plot shows the global part
of the left jet mass, i.e.\ the result without including $S_{\rm
  NG}(\mu_s,\mu_h)$. The difference between the two curves
demonstrates that the non-global pieces have an important effect on
the distribution.  Note that the distributions shown in the plot are
obtained from taking the derivative of the resummed cumulant $R(\rho_L)$ in \eqref{eq:cumulL} with respect to $\rho_L$. For fixed
scales, integrating and differentiating would commute, but we choose
the values of the scales in the cumulant and then take the derivative,
which is advantageous, as explained in \cite{Almeida:2014uva}. One
benefit is that the spectrum is automatically normalized since
$R(\rho_L) \to 1$ for $\rho_L=1$ (the true upper limit of the spectrum
is at a lower value and one often modifies the resummation
prescription such that the result vanishes beyond the kinematical
limit; for simplicity we will not do this here).

\begin{figure}[t!]
\begin{center}
\begin{tabular}{ccc}
\includegraphics[width=0.45\textwidth]{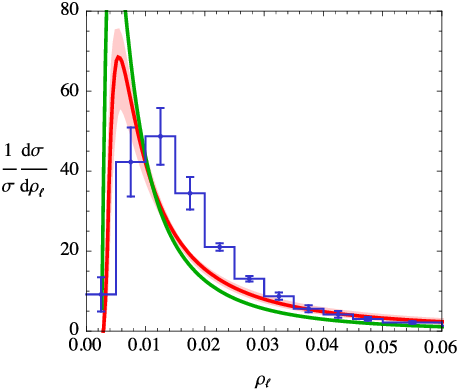} && \includegraphics[width=0.45\textwidth]{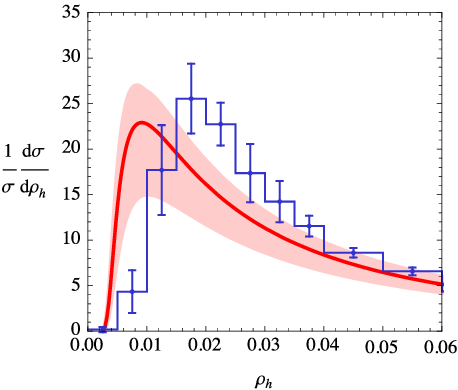} 
\end{tabular}
\end{center}
\vspace{-0.3cm}
\caption{The red bands show the NLL result for the light-jet mass (left) and the heavy-jet mass (right), compared to {\sc Aleph} data (blue) \cite{Buskulic:1992hq}.  The green line is the purely global part of the light-jet mass distribution and peaks at a value of about 110. \label{fig:NLLheavylight}}
\end{figure}

Our plots also include experimental results from the {\sc Aleph}
collaboration \cite{Buskulic:1992hq}. The LEP experiments have
measured the light-jet and heavy-jet mass distributions and we have
used relation \eqref{eq:lightjet} to convert their measurements into a
result for the left-jet mass, naively adding the uncertainties on the
two distributions in quadrature. It is obviously better to directly
compare to the experimental result for the individual measurements,
which is done in Figure \ref{fig:NLLheavylight}. The comparison
  shows that non-perturbative effects, which will shift the peak to
  the right, are important at low values of $\rho_L$, where the
  distribution is large. This is expected since the soft scale is
  $\mu_s \sim \rho_L Q$ and takes non-perturbative values near the
  peak, especially for the light-jet mass. To reproduce the data, one
would have to include such non-perturbative effects, and should also match to the
fixed-order results to get a better description at higher values of
$\rho_\ell$ and $\rho_h$. For the moment, we will not pursue these
issues further. Our goal was to assess whether non-global effects are
phenomenologically relevant and our results clearly show that this is
indeed the case for the non-global hemisphere event shapes.

\section{Conclusions and outlook\label{sec:conclusions}}

We have studied the factorization of large logarithmic corrections appearing in non-global
hemisphere-mass observables at $e^+ e^-$ colliders.  We focused our analysis on two particular cases: {\it i.})
the double differential cross section with respect to the left and right hemisphere 
masses $M_L$ and $M_R$ in the limit where $M_L \ll M_R \ll Q$, and   
{\it ii.}) the left-jet mass distribution in the limit  where $M_L\ll Q$.   Our main result in the first case was the
derivation of a factorization formula for the hemisphere soft function $S(\omega_L,\omega_R)$ in the 
limit $\omega_L \ll \omega_R$, while in the second case we presented a novel factorization formula for the 
differential cross section itself. 

While the specifics of the two cases are slightly different, the ideas 
behind them are rather general, and indeed for the most part could be adapted from the 
analysis of cone-jet cross sections in \cite{Becher:2016mmh}.  In particular, the key feature of 
factorization formulas for such non-global observables is that additional wide-angle emissions of hard partons 
at each order in perturbation theory build up  a tower of multi-Wilson-line operators in the effective field theory.
The matrix elements of these operators define multi-Wilson-line soft functions, which appear 
in angular convolution integrals with their (distribution valued) Wilson coefficients, referred to as multi-parton
hard functions.  

We confirmed the validity of our factorization formulas through
explicit NNLO calculations.  For the hemisphere soft function, we
showed that our results reproduce the known analytic ones from
\cite{Kelley:2011ng, Hornig:2011iu}, including all constant and
logarithmic pieces appearing in the limit $\omega_L\ll \omega_R$.  For
the light-jet mass, on the other hand, we obtained only the
logarithmically enhanced NNLO corrections, and validated them through
numerical comparisons with event generators.  In both cases, the main
new perturbative results presented here were those for the
multi-parton hard functions, since other ingredients appearing in the
factorization could be taken from the literature. We calculated these
to NLO in the case of the left-jet mass, and to NNLO in the case of
the hemisphere soft function, thus providing a non-trivial example at
NNLO of the renormalization procedure involving mixing of
multi-Wilson-line operators characteristic of non-global observables.

The factorization formulas derived here provide the basis for all-orders resummation of non-global logarithms for these observables. To get an idea of the size of the effects, we have used the known result for the leading non-global logarithms in the hemisphere soft function to obtain the left-jet mass distribution at NLL. We find that the non-global effects, evaluated in the large-$N_c$ limit, are of the same magnitude as other NLL effects. For precision predictions of non-global observables, it would be important to include also higher-logarithmic effects. The necessary ingredients are available: we have computed the one-loop soft functions and hard functions and the relevant two-loop anomalous dimensions can be extracted from the work of \cite{Caron-Huot:2015bja}. Since one has to exponentiate an infinite-dimensional anomalous dimension matrix, it is not possible to obtain analytic results and the resummation has to be performed numerically. One approach is to incorporate the corrections into the parton-shower framework used to compute the leading logarithmic corrections. It will be interesting to analyze how this can be done in an efficient way and to use our framework to produce precision predictions for non-global observables.

\begin{acknowledgments}	
We thank Pier Francesco Monni, Matthias Neubert and Lorena Rothen for discussions and Zoltan Trócsányi for sending us results for the light-jet mass obtained using the CoLoRFulNNLO framework. The research of T.B.\ is supported by the Swiss National Science Foundation (SNF) under grant 200020\_165786. The authors thank the ESI Vienna for hospitality and support during early stages of this work.
\end{acknowledgments}   

\appendix

\section{\boldmath Absence of leading-power collinear contributions to $S(\omega_L,\omega_R)$ \unboldmath}\label{sec:Jetfun}

One might expect that left-collinear modes with scaling
\begin{align}
(n\cdot p_c, \bar{n}\cdot p_c,  p_c^\perp) \sim (1,\kappa,\sqrt{\kappa})\, \omega_R 
\end{align}
could contribute to the hemisphere soft function, since they have $\bar{n}\cdot p\sim \omega_L$, as required.
The operator definition for the associated leading-power jet function has the form
\begin{align}
J_c(\omega_L) = & \sum_{X_L} \left|  \langle X_L | W_{\bar n}^\dag W_n |0 \rangle \right|^2  \delta(\omega_L - \sum_i \bar{n}\cdot P_L^i) \,,
\end{align}
where the Wilson lines $W_n$ are built from collinear fields and are invariant under rescaling of the reference
vector. The multipole expansion ensures that the left-collinear fields are always in the left hemisphere and for this reason, the collinear particles do not contribute to $\omega_R$.
According to its definition the jet function transforms as $J \to J/\alpha$ under the transformation
\begin{align}
\bar{n} &\to \alpha\, \bar{n}\, , &\omega_L &\to \alpha \, \omega_L \,, 
\end{align}
or equivalently 
\begin{align}\label{eq:scaling}
J(\alpha\, \omega_L) =  \frac{1}{\alpha} J(\omega_L)\,.
\end{align}
The $n$-loop corrections to $J(\alpha\,\omega_L)$ scale as $\omega_L^{-1-2 n \epsilon}$ and are thus incompatible with this scaling relation \eqref{eq:scaling}. We conclude that they must all vanish so that $J(\omega_L)=\delta(\omega_L)$ to all orders. The leading-power jet function is thus trivial and can be omitted. We note that power corrections do involve nontrivial collinear contributions, as can be checked through an explicit computation of the 
hemisphere soft function using the method of regions.

\section{Bare ingredients for the hemisphere soft function\label{sec:Barefun}}

In the main text, we have mostly presented renormalized results and have reconstructed the renormalized  hemisphere soft function by combining renormalized ingredients. For completeness, we list here also the bare functions. These can be extracted from the results in \cite{Kelley:2011ng, Hornig:2011iu} and they are used in Appendix \ref{sec:renorm} to derive the renormalized expressions. 

The renormalization is interesting from an effective theory point of view and key to perform the resummation. However, to obtain the fixed-order result one can also combine the bare ingredients given in this appendix to recover the bare hemisphere soft function. The bare ingredients are also what is obtained when performing the method of regions computation. At NNLO, the regions computation yields three terms: i) purely hard contributions, ii) purely soft ones, and iii) a mixed contribution with one hard gluon on the right and a soft one on the left. We now list these in turn.

Let us first give the result for the purely hard corrections. They consist of a double-real emission part and a virtual correction to single gluon emission. In the effective theory language they are
\begin{align}
\big\langle \bm{\mathcal{\widetilde{H}}}^{S(2)}_1(\hat\theta_1,\tau_R,\e) \otimes \bm{1} \big\rangle  =& \left(\frac{\mu}{\tau_R}\right)^{4\e}  
C_F C_A v_A,
 \\
 \big\langle \bm{\mathcal{\widetilde{H}}}^{S(2)}_2(\{\underline{n}\},\tau_R,\e) \otimes \bm{1} \big\rangle  =&
  \left(\frac{\mu}{\tau_R}\right)^{4\e}  
\left[C_F^2 h_F^2/2 + C_F C_A h_A + C_F T_F n_f h_f \right] .
\end{align}
When integrating also over the angles to compute these terms, one recovers the standard phase-space integration and the evaluation of these contributions simply amounts to computing the corrections to the Wilson line matrix element \eqref{hardAmpS} in which all particles fly into the right hemisphere. This computation was performed in \cite{Kelley:2011ng, Hornig:2011iu} and we can extract the coefficients $h_F$, $h_A$, $h_f$ and $v_A$ from those papers. The results are 
\begin{align}
h_F = &  -\frac{2}{\e^2} - \frac{\pi^2}{2} - \frac{14\zeta_3}{3} \e - \frac{7\pi^4}{48}\e^2   \nno  \, ,\\
h_A = &  -\frac{1}{\e^4} - \frac{11}{6\e^3} + \frac{1}{\e^2} \left( - \frac{67}{18} - \pi^2\right) + \frac{1}{\e} \left( - \frac{193}{27} - \frac{11\pi^2}{4} - \frac{35\zeta_3}{3} \right) - 
\frac{1196}{81} - \frac{67\pi^2}{12} \nno \\
&  - \frac{473\zeta_3}{9}  - \frac{31\pi^4}{40} \,,
\nonumber \\ 
h_f= & \frac{2}{3\e^3} +\frac{10}{9\e^2}+\frac{1}{\epsilon}\left(\frac{38}{27}+\pi^2  \right)
+\frac{238}{81}+\frac{5\pi^2}{3}+\frac{172\,\zeta_3}{9} \,,  \nonumber \\
v_A = & \frac{1}{\e^4} + \frac{5\,\pi^2}{6\, \e^2} + \frac{56\,\zeta_3}{3\, \e}  + \frac{113\, \pi^4}{120}\,.
\end{align}
While the results above are related to hard gluon emissions into the right hemisphere, soft gluons can radiate into either hemisphere.  Therefore, 
unlike at NLO, the result for the NNLO corrections to $\bm{\mathcal{\widetilde{S}}}_1$ are not simply related
to the hard gluon emissions.  In fact, one has 
\begin{align}
\bm{\mathcal{\widetilde{S}}}_1^{(2)}(\tau_L,\e) =
\left(\frac{\mu}{\tau_L}\right)^{4\e}  
\left[C_F^2 h_F^2/2+ C_F C_A s_A + C_F T_F n_f s_f \right] ,
\end{align}
with 
\begin{align}
s_A- h_A - v_A = &\frac{1}{\epsilon}\left(-\frac{2}{3} +\frac{22\pi^2}{9}-4\zeta_3 \right) 
+\frac{40}{9}-\frac{134\pi^2}{27}+\frac{8\pi^4}{45}+\frac{44\zeta_3}{3}  \,, 
\nonumber \\
s_f- h_f = & \frac{1}{\epsilon}\left(\frac{4}{3} -\frac{8\pi^2}{9}\right) 
-\frac{68}{9}-\frac{16\zeta_3}{3}  +\frac{64\pi^2}{27} \,.
\end{align}
The differences quoted above are due to opposite-side emissions only and they
contribute to subleading NGLs. These opposite-side contributions can be obtained from the computations in \cite{Kelley:2011ng, Hornig:2011iu} by sending the right hemisphere energy $\omega_R$ to infinity because $\omega_R$ is much larger than the momentum components of the soft radiation. We have verified that a direct computation of the corresponding diagrams gives the same result.

The final NNLO contribution is the convolution of NLO terms:
\begin{align}
\langle  \bm{\mathcal{\widetilde{H}}}^{S(1)}_1 \otimes \bm{\mathcal{\widetilde{S}}}^{(1)}_2 \rangle(\tau_L,\tau_R,\epsilon) &  =
 \left(\frac{\mu}{\tau_L} \right)^{2\epsilon} \left(\frac{\mu}{\tau_R}\right)^{2\epsilon} \left(  C_F^2 \, p_F + C_F C_A  \, p_A  \right)  , 
 \end{align}
 where 
 \begin{align}
 p_A =&  \frac{2\pi^2}{3\epsilon^2} +\frac{4\zeta_3}{\epsilon}  +\frac{29\pi^4}{45}  \,, \nno\\
 p_F = & \frac{4}{\e^4} + \frac{2\pi^2}{\e^2} + \frac{56\zeta_3}{3\e} + \frac{5\pi^4}{6} \, .
 \end{align}
It is worth noting that the product coefficient $p_A$ induced through the convolution of NLO functions
is reproduced by the regions calculation of opposite-side gluon contributions, one with a hard scaling and the other with a soft scaling.  This type of contribution is responsible for the leading NGLs, as well as part of the subleading ones.

Evaluating the full NNLO expression according to (\ref{eq:NNLOfact}) then yields
\begin{align}
\label{eq:s2bare}
\widetilde{s}^{(2)}(\tau_L,\tau_R,\epsilon)&=
\left[\left(\frac{\mu}{\tau_L}\right)^{4\e}  + \left(\frac{\mu}{\tau_R}\right)^{4\e} \right]
\left[C_F^2 h_F^2 /2 + C_F C_A (h_A+v_A) + C_F T_F n_f h_f \right] \nonumber \\
&
+\left(\frac{\mu}{\tau_L}\right)^{4\e}
\left[ C_F C_A ( s_A-h_A-v_A) + C_F T_F n_f (s_f- h_f) \right] \nno \\
& + \left(\frac{\mu}{\tau_L}\right)^{2\e} \left(\frac{\mu}{\tau_R}\right)^{2\e}
\left[ C_F C_A p_A+C_F^2 p_F \right]  
\end{align}
where the same-side contributions are in the first line, and the opposite-side contributions in the second and third.

To obtain the renormalized function, we need to multiply by the renormalization factor $\tilde{Z}_S$ introduced in \eqref{eq:softren}. Given the product structure of the factorization theorem \eqref{eq:factMLMR} in Laplace space, it must have a factorized form
\begin{align}
\label{eq:HemisphereZ}
\tilde{Z}_S(\tau_L, \tau_R ,\epsilon, \mu) =\tilde{z}_s(\tau_L, \epsilon, \mu)\tilde{z}_s(\tau_R, \epsilon, \mu) \, ,
\end{align}
where  $\tilde{z}_s$ satisfies the RG equation 
\begin{align}
\frac{d}{d\ln\mu}\tilde{z}_s(\tau, \epsilon, \mu) &= \left[ 2\Gamma_{\rm cusp}\ln\left(\frac{\tau}{\mu}\right) + \gamma_S \right] \tilde{z}_s(\tau, \epsilon, \mu) \,.
\end{align}
Solving this equation perturbatively gives
\begin{align}
\ln(\tilde{z}_s(\tau, \epsilon, \mu) )&=\frac{\alpha_s}{4\pi}\left[\frac{\Gamma_0}{2\e^2}
-\frac{1}{\epsilon}\left(\Gamma_0 L + \frac{\gamma^S_{0}}{2}\right)
\right] \nonumber \\
& + \left(\frac{\alpha_s}{4\pi} \right)^2\left[ -\frac{3\beta_0\Gamma_0}{8\e^3}
+\frac{\Gamma_1}{8\e^2}+\frac{1}{2\e^2}\left(\Gamma_0 L+ \frac{\gamma^S_{0}}{2}\right)\beta_0
-\frac{1}{2\e}\left(\Gamma_1 L+ \frac{\gamma^S_{1}}{2}\right) 
  \right] ,
\end{align}
where $L=\ln(\tau/\mu)$. For convenience we give the necessary anomalous dimension in the above expression. The expansion of the anomalous dimensions in the strong coupling constant reads
\begin{align}
\Gamma_{\rm cusp} = & \sum_{n=0}^{\infty}\left(\frac{\as}{4\pi}\right)^{n+1} \Gamma_{n}\,, \nno &
\gamma_{ S} = & \sum_{n=0}^{\infty}\left(\frac{\as}{4\pi}\right)^{n+1} \gamma^{S}_{n}\,,
\end{align}
with
\begin{align}
\Gamma_{0} = & \, 4 C_F,  & \Gamma_{1} = & \left( \frac{268}{9} - \frac{4\pi^2}{3} \right) C_F C_A - \frac{80}{9} C_F T_F n_f,
 \end{align}
 and
\begin{align} 
\gamma^{S}_{0}= & \, 0, & \gamma^{S}_{1}=&  \left( -\frac{808}{27} + \frac{11\pi^2}{9} + 28 \zeta_3 \right) C_F C_A + \left( \frac{224}{27} - \frac{4\pi^2}{9} \right) C_F T_F n_f.
\end{align}
To perform the NLL resummation in Section \ref{sec:NLL} we also need the anomalous dimensions
\begin{align}
\gamma^{J}_0 &= -3 C_F\,,& 
\beta_0 &= \frac{11}{3}\,C_A - \frac43\,T_F n_f \,, &  \beta_1 &= \frac{34}{3}\,C_A^2 - \frac{20}{3}\,C_A T_F n_f
    - 4 C_F T_F n_f \,.
\end{align}

\section{NNLO renormalization for the factorized hemisphere soft function \label{sec:renorm}}

We have presented the renormalization equations for the component hard and soft functions entering
the factorization formula for the hemisphere soft function in \eqref{eq:HmZfac} and 
\eqref{eq:SmZfac}. Using that $\Htilde{0}^{S(1)}=\bm{1}$ and writing out these  equations explicitly to NNLO and suppressing all functional dependence 
except for $\mu$ and $\epsilon$ on the right-hand side, we find the relations
\begin{align}
\Htilde{1}^{S(1)}(\{\underline{n}\},\tau_R,\mu)&=\Htilde{1}^{S(1)}(\epsilon) 
- \Ztilde{01}^{(1)}(\epsilon,\mu) \,, \nonumber \\
\Htilde{1}^{S(2)}(\{\underline{n}\},\tau_R,\mu)&=
\Htilde{1}^{S(2)}(\epsilon) 
-\Ztilde{01}^{(2)}(\epsilon,\mu) + \Ztilde{01}^{(1)}(\epsilon,\mu) \Ztilde{11}^{(1)}(\epsilon,\mu)
 \nonumber \\&
 -\Htilde{1}^{S(1)}(\epsilon)\left[\Ztilde{11}^{(1)}(\epsilon,\mu) +\frac{\beta_0}{\epsilon}\right]  ,
\nonumber \\
\Htilde{2}^{S(2)}(\{\underline{n}\},\tau_R,\mu)&=
\Htilde{2}^{S(2)}(\epsilon) 
-\Ztilde{02}^{(2)}(\epsilon,\mu) + \Ztilde{01}^{(1)}(\epsilon,\mu) \Ztilde{12}^{(1)}(\epsilon,\mu)
-\Htilde{1}^{S(1)}(\epsilon) \Ztilde{12}^{(1)}(\epsilon,\mu)  \,,
\end{align}
where the term involving $\beta_0$ arises because the bare functions were expanded in the bare coupling instead of the renormalized one. Similarly, for the soft function we find
\begin{align}
\Stilde{1}^{(1)}(\{\underline{n}\},\tau_L,\mu)=& \left[\tilde{Z}_S(\epsilon,\mu) \Stilde{1}(\epsilon)\right]^{(1)}
+  \Ztilde{01}^{(1)}(\epsilon,\mu) \hat\otimes \bm{1} \, , \nonumber \\
\Stilde{1}^{(2)}(\{\underline{n}\},\tau_L,\mu)= &\left[\tilde{Z}_S(\epsilon,\mu) \Stilde{1}(\epsilon)\right]^{(2)}
+\Ztilde{01}^{(1)} \hat\otimes \left[\tilde{Z}_S(\epsilon,\mu) \Stilde{2}(\epsilon)\right]^{(1)} 
\nonumber \\ &
+\left(\Ztilde{01}^{(2)}(\epsilon,\mu) +\Ztilde{02}^{(2)}(\epsilon,\mu) \right) \hat\otimes {\bm 1} 
\, , \nonumber \\
\Stilde{2}^{(1)}(\{\underline{n}\},\tau_L,\mu) =&
 \left[\tilde{Z}_S(\epsilon,\mu) \Stilde{2}(\epsilon)\right]^{(1)}
 +\Ztilde{11}^{(1)}(\epsilon,\mu)+ \Ztilde{12}^{(1)}(\epsilon,\mu) \hat{\otimes} \bm{1} \,.
\end{align} 
Here  $\left[ \dots \right]^{(2)}$ and $\Ztilde{lm}^{(2)}$ refer to the second-order coefficients in the renormalized coupling, while $\Stilde{1}^{(2)}(\epsilon)$ denotes the second order coefficient of the bare coupling.
Notice that $\Stilde{2}^{(1)}$ is a regular function in its arguments, so the equations above imply
that  $\Ztilde{12}^{(1)}$ and $\Ztilde{11}^{(1)}$ are also regular functions and not distributions.
It follows that the renormalized NLO functions are simply obtained from the bare 
functions by dropping the poles.  Moreover, the following linear combinations of renormalization
factors are immediately obtained
\begin{align}
\label{eq:Zcombs}
& \langle \Ztilde{11}^{(1)}(\epsilon,\mu)+ \Ztilde{12}^{(1)}(\epsilon,\mu) \hat{\otimes} \bm{1} \rangle
=\, C_F\left[-\frac{2}{\e^2}+\frac{4}{\e} L_R \right]
-2 \, C_A \frac{ \ln(1-\hat{\theta}_1^2)}{\epsilon} \,, \nonumber \\ 
 &\big\langle \big[\Ztilde{01}^{(2)}(\epsilon,\mu) +\Ztilde{02}^{(2)}(\epsilon,\mu) \big] \hat{\otimes} {\bm 1} \big\rangle= C_F^2 \bigg[  \frac{2}{\e^4} - \frac{8}{\e^3} L_R + \frac{8}{\e^2} L_R^2   \bigg] \nno \\
 & + 
 C_A C_F \bigg[\frac{11}{2\e^3}-\frac{1}{\e^2}\left(\frac{67}{18}+\frac{\pi^2}{6} +\frac{22}{3}L_R \right) + \frac{1}{\epsilon}\bigg(-\frac{193}{27}-\frac{11\pi^2}{12}+3\zeta_3 
  +  \left(\frac{134}{9}-\frac{2\pi^2}{3}\right)L_R\bigg)\bigg] \nonumber \\
& + C_F T_F n_f\bigg[ -\frac{2}{\epsilon^3}+\frac{1}{\e^2}\left(\frac{10}{9}+\frac{8}{3}L_R\right)    
+\frac{1}{\epsilon}\left(\frac{38}{27}+\frac{\pi^2}{3}-\frac{40}{9}L_R\right)          \bigg]\, .
\end{align}

For the renormalized soft function we obtain the result in
\eqref{eq:softOne}.  Because only the linear combinations of
renormalization factors listed in (\ref{eq:Zcombs}) above is
determined, and because we have the bare functions only after
integrating over angles, we can only determine the combination
$\langle \Htilde{1}^{S(2)}(\{\underline{n}\},\tau_R,\mu)\otimes \bm {1} +
\Htilde{2}^{S(2)}(\{\underline{n}\},\tau_R,\mu)\otimes \bm {1}\rangle$ of
NNLO hard functions. The result for this combination was given in
\eqref{eq:hard2}.

\section{Bare ingredients for the light-jet mass\label{sec:bareLjet}}

In the main text, we provided the ingredients to obtain the light-jet mass distribution from renormalized quantities, but equally well one can construct the result starting from their bare counterparts. To this end, we collect here all the two-loop bare ingredients for the light-jet mass case. The bare hard function $H(Q,\e)$ and soft function $\bm{\mathcal{\widetilde{S}}}_1(\tau,\e)$ have been given in Appendix A of \cite{Becher:2016mmh} and we only list the new two-loop ingredients. The first is one-loop bare hard function $\bm{\mathcal{H}}_2^{(1)}$ convoluted with the trivial leading-order soft function
\begin{align}
&\sum_{i=q,\bar{q}}\langle \bm{\mathcal{H}}_2^{i, (1)} \otimes \bm{1} \rangle = C_F \sigma_0 \left(\frac{\mu}{Q}\right)^{2\e} \Bigg[ \frac{2}{\e^2} + \frac{3}{\e}  +\frac{29}{3}-\frac{3 \pi ^2}{2}-2 \ln
   ^2 2+\frac{5 \ln 3}{4} -4 \, \text{Li}_2\left(-\frac{1}{2}\right)  \nno \\ 
   & \hspace{3.8cm} + \e\Bigg(\frac{169}{6}  -\frac{11 \pi ^2}{6} -\frac{76 \zeta_3}{3} +\frac{32 \ln ^3 2}{3}-\frac{7 \ln
   ^2 2}{4}-18 \ln ^2 2 \ln 3+\frac{15 \ln ^2 3}{8} \nno \\
   & \hspace{3.8cm} +6 \ln 2 \ln
   ^2 3-\frac{4}{3} \pi ^2 \ln 2+\frac{39 \ln 3}{8}  -\frac{7 }{2} \, \text{Li}_2\left(-\frac{1}{2}\right) -12
   \text{Li}_2\left(-\frac{1}{2}\right) \ln 3 \nno \\
   & \hspace{3.8cm} -16 \,
   \text{Li}_3\left(-\frac{1}{2}\right)-6\,  \text{Li}_3\left(\frac{3}{4}\right)  \Bigg)  + \mathcal{O}(\e^2) \Bigg], \nno\\
& \langle \bm{\mathcal{H}}_2^{g, (1)} \otimes \bm{1} \rangle =C_F \sigma_0 \left(\frac{\mu}{Q}\right)^{2\e} \Bigg[ -\frac{1}{6}+\frac{\pi ^2}{3}+2 \ln^2 2-\frac{5 \ln 3}{4} +  4\, \text{Li}_2\left(-\frac{1}{2}\right)  + \e \Bigg( -\frac{11}{12}+\frac{\pi ^2}{12} \nno \\
& \hspace{3.8cm} +\frac{22 \zeta_3}{3} -\frac{8 \ln ^3 2}{3}+\frac{4 \ln
   ^3 3}{3}+\frac{7 \ln ^2 2}{4}+10 \ln ^2 2 \ln 3-\frac{15 \ln ^2 3}{8}  \nno \\
   &\hspace{3.8cm} -6
   \ln 2 \ln ^2 3 +\frac{4}{3} \pi ^2 \ln 2-\frac{39 \ln 3}{8} + \frac{7}{2} \, \text{Li}_2\left(-\frac{1}{2}\right) +12 \, 
   \text{Li}_2\left(-\frac{1}{2}\right) \ln 3 \nno \\
   &\hspace{3.8cm} -8 \, 
   \text{Li}_3\left(\frac{1}{3}\right)  +2 \, \text{Li}_3\left(\frac{3}{4}\right) \Bigg) + \mathcal{O}(\e^2) \Bigg]\, .
\end{align}
Each of the results includes transcendental numbers other than $\zeta$-values, but they exactly cancel out in the sum of both contributions. For completeness we also list the bare jet functions in Laplace space. The two-loop quark jet function reads
\begin{align}
\tilde j_{q,{\rm bare}}(\tau Q,\e) = \, & 1 + \frac{\alpha_0 C_F}{4\pi}\,\left( \frac{\mu^2}{\tau Q} \right)^\e \Bigg[ \frac{4}{\e^2} + \frac{3}{\e} + 7 - \frac{2\pi^2}{3}  + \e\left(14 - \frac{\pi^2}{2} - 8\zeta_3 \right) \nno \\
& \hspace{-2cm} + \e^2 \left( 28 - \frac{7\pi^2}{6} - 6\zeta_3 - \frac{\pi^4}{10}\right)\Bigg] +  \left( \frac{\alpha_0}{4\pi} \right)^2 \left( \frac{\mu^2}{\tau Q} \right)^{2\e} 
    \left(  C_F^2 j_F + C_F C_A j_A + C_F T_F n_f j_f \right),
\end{align}
with
\begin{align}
j_F &= \frac{8}{\e^4} + \frac{12}{\e^3} + \frac{1}{\e^2} \left( \frac{65}{2} - \frac{8\pi^2}{3} \right) + \frac{1}{\e}  \left( \frac{311}{4} - 5\pi^2 - 20\zeta_3 \right) +\frac{1437}{8}-\frac{57 \pi ^2}{4} -54 \zeta_3 +\frac{5 \pi ^4}{18} , \nno \\
j_A &= \frac{11}{3\e^3} + \frac{1}{\e^2} \left( \frac{233}{18} - \frac{\pi^2}{3} \right) + \frac{1}{\e} \left( \frac{4541}{108} - \frac{11\pi^2}{6}- 20\zeta_3  \right)  +\frac{86393}{648}-\frac{221 \pi ^2}{36} -\frac{142 \zeta_3}{3} -\frac{37 \pi ^4}{180} , \nno \\
j_f & = - \frac{4}{3\e^3} - \frac{38}{9\e^2} + \frac{1}{\e} \left( - \frac{373}{27} + \frac{2\pi^2}{3}\right)-\frac{7081}{162}+\frac{19 \pi ^2}{9} + \frac{32 \zeta_3}{3}, 
\end{align}
and the one-loop gluon result has the form
\begin{align}
 \tilde j_{g,{\rm bare}}(\tau Q,\e) = & \, 1 + \frac{\alpha_0}{4\pi}\,\left( \frac{\mu^2}{\tau Q} \right)^\e 
\Bigg[ C_A \Bigg(\frac{4}{\e^2} + \frac{11}{3\e} + \frac{67}{9} - \frac{2\pi^2}{3} \Bigg) +  T_F n_f \Bigg( - \frac{4}{3\e} - \frac{20}{9}\Bigg)   \Bigg] \,. 
\end{align}
Next, we consider the convolution of the one-loop hard and soft functions. Since we are only interested in the logarithmic terms in the cross section, it is sufficient to give the divergent parts of the convolution, which have the form
\begin{align}
&\sum_{i=q,\bar{q}}\left\langle \bm{\mathcal{H}}_2^{i(1)}  \otimes \bm{\mathcal{\widetilde{S}}}_2^{(1)}  \right\rangle_{\rm div.}  = \left(\frac{\mu^2}{\tau Q}\right)^{2\e} \sigma_0 \Bigg[C_F^2 \Bigg( -\frac{4}{\e^4} - \frac{6}{\e^3} + \frac{M_{F,\, q}^{[-2]}}{\e^2}  + \frac{M_{F,\, q}^{[-1]}}{\e} \Bigg)  \nno \\
& \hspace{9cm} + C_F C_A\left( \frac{2\pi^2}{3\e^2} + \frac{M_{A,\, q}^{[-1]}}{\e} \right) \Bigg], \nno \\
&\left\langle \bm{\mathcal{H}}_2^{g(1)}  \otimes \bm{\mathcal{\widetilde{S}}}_2^{(1)}  \right\rangle_{\rm div.}  = \left(\frac{\mu^2}{\tau Q}\right)^{2\e} \sigma_0 \Bigg[ C_F^2  \Bigg(  \frac{ M_{F, \, g}^{ [-1]} }{\e} \Bigg)  + C_F C_A\Bigg( \frac{M_{A, \, g}^{ [-2]}}{\e^2}   +\frac{M_{A, \, g}^{ [-1]}}{\e}  \Bigg) \Bigg] \,,
\end{align}
with
\begin{align}
M_{F,\, q}^{[-2]} &= -\frac{58}{3} + 2\pi^2 + 4 \ln^2 2 -\frac{5}{2}\ln 3 + 8\,\text{Li}_2\left(-\frac{1}{2}\right), \nno \\
M_{F,\, q}^{[-1]} &= -\frac{395}{6}+\frac{23 \pi ^2}{4} + \frac{167 \zeta_3}{6} -\frac{92 \ln ^3 2}{3}+\frac{41
   \ln ^2 2}{4}+48 \ln ^2 2 \ln 3-5 \ln ^2 3-16 \ln 2 \ln
   ^2 3 \nno \\
   & \hspace{.4cm}   +\frac{28 \ln 2}{3}+4 \pi ^2 \ln 2-\frac{337 \ln
   3}{12}+\frac{5}{2} \ln 2 \ln 3  +  \frac{73}{2} \, \text{Li}_2\left(-\frac{1}{2}\right)-8 \,  \text{Li}_2\left(-\frac{1}{2}\right)
   \ln 2   \nno \\
   & \hspace{.4cm} +32 \, \text{Li}_2\left(-\frac{1}{2}\right) \ln 3 +32 \,
   \text{Li}_3\left(-\frac{1}{2}\right) +16 \,  \text{Li}_3\left(\frac{3}{4}\right) , \nno \\
M_{A,\, q}^{[-1]} & =  \frac{33}{8} -\frac{7 \pi^2}{2} +76 \zeta_3 +\frac{64 \ln ^3 2}{3}+8 \ln 2 \ln ^2 3+\frac{5
   \ln ^2 3}{2}-24 \ln ^2 2 \ln  3-\frac{33 \ln ^2 2}{2} \nno \\
   & \hspace{.4cm}  +\frac{39}{2}
   \ln 2 \ln 3+\frac{359 \ln 3}{12}-\frac{8}{3} \pi ^2 \ln 2-52
   \ln 2 -65 \,
   \text{Li}_2\left(-\frac{1}{2}\right)-16 \,
   \text{Li}_2\left(-\frac{1}{2}\right) \ln 3 \nno \\
   & \hspace{.4cm} +40 \,
   \text{Li}_2\left(-\frac{1}{2}\right) \ln 2 -8 \, \text{Li}_3\left(\frac{3}{4}\right)+56 \,
   \text{Li}_3\left(-\frac{1}{2}\right)  + 8 \, I_1 \, ,
 \end{align}
 and
 \begin{align}
M_{F,\, g}^{[-1]} &=  -2 -\frac{55 \,  \pi
   ^2}{12}  -\frac{364 \, \zeta_3}{3}  -20 \ln ^3 2-4 \ln 2 \ln ^2 3-\frac{5 \ln ^2 3}{4}+12 \ln ^2 2 \ln 3-\frac{69 \ln ^2 2}{4} \nno \\
   & \hspace{.4cm} +\frac{51}{2} \ln 2 \ln 3-\frac{23 \ln
   3}{6}+\frac{16}{3} \pi ^2 \ln 2 + 20 \ln 2  - \frac{133}{2} \,  \text{Li}_2\left(-\frac{1}{2}\right)+8 \, 
   \text{Li}_2\left(-\frac{1}{2}\right) \ln 3 \nno \\
   & \hspace{.4cm} - 88 \,
   \text{Li}_2\left(-\frac{1}{2}\right) \ln 2  + 4 \, \text{Li}_3\left(\frac{3}{4}\right)-176 \, 
   \text{Li}_3\left(-\frac{1}{2}\right) + 8 \, I_2\, , \nno \\
M_{A, \, g}^{ [-2]} & = \frac{1}{3}-\frac{2 \pi ^2}{3}-4 \ln
   ^2 2+\frac{5 \ln 3}{2} -8 \, \text{Li}_2\left(-\frac{1}{2}\right), \nno \\
M_{A, \, g}^{ [-1]} & =  \frac{13}{3}+\pi ^2+47 \zeta_3 +36
   \ln ^3 2+\frac{23 \ln ^2 2}{4}-48 \ln ^2 2 \ln 3+5 \ln ^23+16 \ln 2
   \ln ^2 3-\frac{56 \ln 2}{3} \nno \\
   & \hspace{.4cm} -6 \pi ^2 \ln 2+\frac{61 \ln 3}{3}-14 \ln
   2 \ln 3  + \frac{23 }{2}\, \text{Li}_2\left(-\frac{1}{2}\right) +56 \,
   \text{Li}_3\left(-\frac{1}{2}\right)-16 \, \text{Li}_3\left(\frac{3}{4}\right) \nno \\
   & \hspace{.4cm} +48 \,
   \text{Li}_2\left(-\frac{1}{2}\right) \ln 2-32
   \text{Li}_2\left(-\frac{1}{2}\right) \ln 3 \,.
\end{align}
The result for the coefficients involves two angular integrals $I_1$ and $I_2$ which we were not able to evaluate in closed form. They are
\begin{align}
I_1 & = \int_0^{1/\sqrt{3}} d\hat{\theta}_2 \int_{\hat{\theta}_2}^{-\hat{\theta}_2 + \sqrt{1+ \hat{\theta}_2^2}} d \hat{\theta}_1 \,f(\hat{\theta}_1,\hat{\theta}_2) \ln(1-\hat{\theta}_2^2) = -0.0423782819\, , \nno \\
I_2 & = \int_0^{1/\sqrt{3}} d\hat{\theta}_2 \int_{\hat{\theta}_2}^{-\hat{\theta}_2 + \sqrt{1+ \hat{\theta}_2^2}} d \hat{\theta}_1 \,g(\hat{\theta}_1,\hat{\theta}_2) \ln(1-\hat{\theta}_2^2) = -0.0145491799\,,
\end{align}
with 
\begin{align}
f(\hat{\theta}_1,\hat{\theta}_2) &= \frac{ 2 \hat{\theta}_1 \hat{\theta}_2(1-\hat{\theta}_1 ^2)(1-\hat{\theta}_2^2)  + 2\hat{\theta}_2^2(1+\hat{\theta}_1^4 ) + \hat{\theta}_1^2(1-\hat{\theta}_2^2)^2 }{\hat{\theta}_1 (\hat{\theta}_1 + \hat{\theta}_2)^3}\, , \nno \\
g(\hat{\theta}_1,\hat{\theta}_2) &= \frac{\hat{\theta}_1^2(1+\hat{\theta}_2^2)^2 + \hat{\theta}_2^2(1+\hat{\theta}_1^2)^2}{ (\hat{\theta}_1+\hat{\theta}_2)^2 (1-\hat{\theta}_1 \hat{\theta}_2)}\, .
\end{align}
In the main text, we considered in the convolution of the renormalized one-loop hard and soft functions. The form of the convolution was given in \eqref{eq:renConvq} and \eqref{eq:renConvg}. The coefficients of the logarithmic terms in these two formulas are closely related the coefficients of the divergences given above. Explicitly, they read 
\begin{align}\label{eq:McoeffsRen}
M_{q,F}^{(2)} &=-\frac{116}{3}+6 \, \pi ^2+8 \ln ^2 2-5 \ln
   3 + 16 \, \text{Li}_2\left(-\frac{1}{2}\right), \nno \\
   M_{q,F}^{(1)} &=19-\frac{43 \, \pi
   ^2}{6}+ 27 \zeta_3 +\frac{56 \ln ^3 2}{3}-\frac{27 \ln ^2 2}{2}-24 \ln ^2 2 \ln
   3+\frac{5 \ln ^2 3}{2}+8 \ln 2 \ln ^2 3 \nno \\
   & \hspace{.4cm} -\frac{56 \ln 2}{3}-\frac{8}{3} \pi ^2 \ln 2+\frac{110 \ln 3}{3}-5 \ln 2 \ln 3 -59\, \text{Li}_2\left(-\frac{1}{2}\right)-8 \, \text{Li}_3\left(\frac{3}{4}\right) \nno \\
   & \hspace{.4cm} +16 \,
   \text{Li}_2\left(-\frac{1}{2}\right) \ln  2-16 \, 
   \text{Li}_2\left(-\frac{1}{2}\right) \ln 3\, , \nno\\
M_{q,A}^{(1)} & = -\frac{33}{4} +7
   \pi ^2 -136 \zeta_3-\frac{128 \ln ^3 2}{3}-16 \ln  2 \ln ^2 3-5 \ln
   ^2 3+48 \ln ^2 2 \ln  3+33 \ln ^2 2 \nno \\
   & \hspace{.4cm} -39 \ln  2 \ln  3-\frac{359
   \ln  3}{6}+\frac{16}{3} \pi ^2 \ln  2+104 \ln  2   +130 \, 
   \text{Li}_2\left(-\frac{1}{2}\right)+32  \, \text{Li}_2\left(-\frac{1}{2}\right)
   \ln  3  \nno \\
   & \hspace{.4cm}-80  \, \text{Li}_2\left(-\frac{1}{2}\right) \ln  2+16 \, \text{Li}_3\left(\frac{3}{4}\right) - 112 \, \text{Li}_3\left(-\frac{1}{2}\right)  -16 \, I_1\, , \nno \\
 M_{g,F}^{(1)} & = 4 +\frac{55 \pi ^2}{6}  +\frac{728 \zeta_3}{3}+40 \ln ^3 2+8 \ln  2 \ln ^2 3+\frac{5
   \ln ^2 3}{2}-24 \ln ^2 2 \ln  3+\frac{69 \ln ^2 2}{2} \nno \\
   & \hspace{.4cm} -51 \ln  2
   \ln  3+\frac{23 \ln  3}{3}-\frac{32}{3} \pi ^2 \ln  2-40 \ln  2 +133 \,
   \text{Li}_2\left(-\frac{1}{2}\right)-16 \, \text{Li}_2\left(-\frac{1}{2}\right)
   \ln  3 \nno \\
   & \hspace{.4cm} +176  \, \text{Li}_2\left(-\frac{1}{2}\right) \ln  2 -8 \, \text{Li}_3\left(\frac{3}{4}\right)+352 \,
   \text{Li}_3\left(-\frac{1}{2}\right) -16\,I_2 \,, \nno \\  
     M_{g,A}^{ (2)} &= \frac{2}{3}-\frac{4 \pi ^2}{3}-8 \ln
   ^2 2 +5 \ln 3 -16 \, \text{Li}_2\left(-\frac{1}{2}\right) , \nno \\
M_{g,A}^{(1)} &=  -5-\frac{7 \pi ^2}{3} -\frac{386 \, \zeta_3}{3} -\frac{88 \ln ^3 2}{3}- \frac{37 \ln ^2 2}{2}+24 \ln
   ^2 2 \ln 3- \frac{5 \ln ^2 3}{2}- 8 \ln 2 \ln ^2 3 \nno \\
   & \hspace{.4cm} +\frac{112 \ln
    2}{3}+\frac{20}{3} \pi ^2 \ln  2-\frac{127 \ln 3}{6}+28 \ln 2 \ln
   3  - 37 \, \text{Li}_2\left(-\frac{1}{2}\right)- 176 \, 
   \text{Li}_3\left(-\frac{1}{2}\right) \nno \\
   & \hspace{.4cm} + 8 \, \text{Li}_3\left(\frac{3}{4}\right)- 96 \,
   \text{Li}_2\left(-\frac{1}{2}\right) \ln 2+16\,
   \text{Li}_2\left(-\frac{1}{2}\right) \ln 3\, .
\end{align}
Since we are only interested in the logarithmic terms in the cross section, we do not list the results for the constants $M_{X}^{(0)}$.

Finally, the divergent part of two-loop hard functions
$\bm{\mathcal{H}}_2^{(2)}$ and
$\bm{\mathcal{H}}_3^{(2)} $ can be inferred from the
requirement that the cross section is finite. The finiteness condition
implies that the divergences are given by
\begin{align}
&\sum_{i=q,\bar{q}} \langle \bm{\mathcal{H}}_2^{i, (2)} \otimes \bm{1}  + \bm{\mathcal{H}}_3^{i, (2)} \otimes \bm{1}  \rangle_{\rm div.}  = \left(\frac{\mu}{Q}\right)^{4\e} \sigma_0\, \Big( C_F^2 H_{F,q}  + C_F C_A  H_{A, q} + C_F T_F n_f  H_{f, q}   \Big)  , \nno \\
 &\langle \bm{\mathcal{H}}_2^{g, (2)} \otimes \bm{1}  + \bm{\mathcal{H}}_3^{g, (2)} \otimes \bm{1}  \rangle_{\rm div.} = \left(\frac{\mu}{Q}\right)^{4\e} \sigma_0\, \Big( C_F^2 H_{F,g} + C_F C_A H_{A,g} \Big) \, .
\end{align}
The second-order coefficients of the different color structures
 for the quark and gluon contributions read
\begin{align}
H_{F ,q} = &  -\frac{6}{\e^4} - \frac{18}{\e^3} + \frac{1}{\e^2} \Bigg[ -\frac{389}{6}+\frac{23 \pi ^2}{3}+4 \ln
   ^2 2-\frac{5 \ln 3}{2}  + 8 \,  \text{Li}_2\left(-\frac{1}{2}\right) \Bigg] + \frac{1}{\e} \Bigg[  -\frac{2245}{12} \nno \\
   & +\frac{211 \pi ^2}{12} +\frac{569 \zeta_3}{6} -12 \ln ^32+\frac{11 \ln
   ^2 2}{4}+24 \ln ^2 2 \ln  3-\frac{5 \ln ^2 3}{2}-8 \ln  2 \ln
   ^2 3-\frac{28 \ln  2}{3} \nno \\
   & +\frac{4}{3} \pi ^2 \ln  2+\frac{29 \ln
    3}{6}-\frac{5}{2} \ln  2 \ln  3   -\frac{21 }{2}\,\text{Li}_2\left(-\frac{1}{2}\right) +8 \, 
   \text{Li}_2\left(-\frac{1}{2}\right) \ln  2 \nno \\
   & +16 \, 
   \text{Li}_2\left(-\frac{1}{2}\right) \ln  3 +32 \, 
   \text{Li}_3\left(-\frac{1}{2}\right)+8 \, \text{Li}_3\left(\frac{3}{4}\right) \Bigg]\, , \nno  \\
  H_{A, q} = & \, \frac{11}{6\e^3} + \frac{1}{\e^2} \left( \frac{83}{9} - 
   \frac{\pi^2}{2} \right)  + \frac{1}{\e} \Bigg[ \frac{8759}{216} -\frac{83 \pi
   ^2}{36} -85 \zeta_3 -\frac{64 \ln ^3 2}{3}-8 \ln  2 \ln
   ^2 3-\frac{5 \ln ^2 3}{2} \nno \\
   & +24 \ln ^2 2 \ln  3+\frac{55 \ln
   ^2 2}{6}-\frac{39}{2} \ln  2 \ln  3-\frac{76 \ln  3}{3}+\frac{8}{3}
   \pi ^2 \ln  2+52 \ln  2 +8 \, \text{Li}_3\left(\frac{3}{4}\right) \nno \\
   & -56 \,
   \text{Li}_3\left(-\frac{1}{2}\right)+\frac{151 
   }{3} \text{Li}_2\left(-\frac{1}{2}\right) +16 \,
   \text{Li}_2\left(-\frac{1}{2}\right) \ln  3-40 \,
   \text{Li}_2\left(-\frac{1}{2}\right) \ln  2 -8 \, I_1 \Bigg] \, , \nno \\
 H_{f,q}  = & - \frac{2}{3\e^3} - \frac{28}{9\e^2} + \frac{1}{\e} \Bigg[  -\frac{431}{27}+\frac{19 \pi
   ^2}{9}+\frac{8 \ln^2 2}{3}-\frac{5 \ln 3}{3} + \frac{16}{3} \, \text{Li}_2\left(-\frac{1}{2}\right) \Bigg]\, ,
\end{align}
\begin{align}
H_{F,g}=& \, \frac{1}{\e} \Bigg [ 2+\frac{55 \pi ^2}{12}+\frac{364 \zeta_3}{3}+20 \ln ^3 2+4 \ln  2 \ln ^2 3+\frac{5
   \ln ^2 3}{4}-12 \ln ^2 2 \ln  3+\frac{69 \ln ^2 2}{4} \nno \\
   & -\frac{51}{2}
   \ln  2 \ln  3+\frac{23 \ln  3}{6}-\frac{16}{3} \pi ^2 \ln  2-20 \ln
    2 +\frac{133
   }{2} \, \text{Li}_2\left(-\frac{1}{2}\right) - 8 \, 
   \text{Li}_2\left(-\frac{1}{2}\right) \ln  3 \nno \\
   & +88\,
   \text{Li}_2\left(-\frac{1}{2}\right) \ln  2 -4 \, \text{Li}_3\left(\frac{3}{4}\right)+176 \, 
   \text{Li}_3\left(-\frac{1}{2}\right) -8 \, I_2 \Bigg]\, , \nno \\
 H_{A,g} = & \,  \frac{1}{\e^2} \Bigg[\frac{1}{3}-\frac{2 \pi ^2}{3}-4 \ln^2 2+\frac{5 \ln 3}{2} -8 \, \text{Li}_2\left(-\frac{1}{2}\right) \Bigg] 
 + \frac{1}{\e} \Bigg[   -\frac{2}{3}-\frac{4 \pi ^2}{3}  -\frac{229 \zeta_3}{3}-\frac{76 \ln ^3 2}{3} \nno \\
 &-\frac{16 \ln
   ^3 3}{3}-\frac{51 \ln ^2 2}{4}+8 \ln ^2 2 \ln  3+\frac{5 \ln
   ^2 3}{2}+8 \ln  2 \ln ^2 3+\frac{56 \ln  2}{3}+\frac{2}{3} \pi ^2 \ln
    2-\frac{5 \ln  3}{6} \nno \\
    & +14 \ln  2 \ln  3  -\frac{51 }{2}\, \text{Li}_2\left(-\frac{1}{2}\right) - 56 \,
   \text{Li}_3\left(-\frac{1}{2}\right)-48 \, \text{Li}_2\left(-\frac{1}{2}\right)
   \ln  2-16 \, \text{Li}_2\left(-\frac{1}{2}\right) \ln  3 \nno \\
   & +32 \, \text{Li}_3\left(\frac{1}{3}\right) + 8 \,
   \text{Li}_3\left(\frac{3}{4}\right)  \Bigg]\, .
\end{align}

\end{document}